\documentclass[11pt,letterpaper]{JHEP3}
\usepackage{epsfig}

\newcommand{\be}{\begin{equation}}
\newcommand{\beqa}{\begin{eqnarray}}
\newcommand{\beq}{\begin{equation}}

\newcommand{\ee}{\end{equation}}
\newcommand{\eeqa}{\end{eqnarray}}
\newcommand{\eeq}{\end{equation}}
\newcommand{\lsim}{\!\mathrel{\hbox{\rlap{\lower.55ex \hbox{$\sim$}} \kern-.34em \raise.4ex \hbox{$<$}}}}
\newcommand{\gsim}{\!\mathrel{\hbox{\rlap{\lower.55ex \hbox{$\sim$}} \kern-.34em \raise.4ex \hbox{$>$}}}}

\newcommand{\ltap}{\raisebox{-.4ex}{\rlap{$\sim$}} \raisebox{.4ex}{$<$}}   % < or ~
\newcommand{\gtap}{\raisebox{-.4ex}{\rlap{$\sim$}} \raisebox{.4ex}{$>$}}   % > or ~\newcommand{\paslash}{\ensuremath \raisebox{0.025cm}{\slash}\hspace{-0.25cm}\partial\/}

\title{Experimental Constraints on Higgs Boson
Decays to TeV-scale Right-Handed Neutrinos}
\author{Michael L. Graesser \\ Department of Physics and NHETC,
 Rutgers University,
Piscataway, NJ 08540\\ Email: \email{mgraesser@physics.rutgers.edu}}

\abstract{The existence of neutrino masses strongly suggests that
right-handed neutrinos exist, but the data do not favor any
particular scale for the Majorana mass parameters. Here I explore
the possibility that these particles exist at the electroweak scale
along with additional new physics at the TeV scale . Higher
dimension operators involving right-handed neutrinos and the Higgs
boson can introduce new decay modes of the Higgs boson,
significantly modifying its phenomenology if it is light. With
minimal flavor violation the Higgs boson cascade decays to 6
particles containing two highly displaced vertices. Each displaced
vertex produces an odd number of leptons, leading to a dramatic
signature of overall lepton violation at each vertex. I discuss the
limits from the Tevatron, and find that they are close to exploring
interesting regions of parameters, while limiting others. Moving
beyond minimal flavor violation, cascade decays of the Higgs boson
into as many as 14 particles can occur.}
\preprint{RUNHETC-04-2007}

\begin{document}

\section{Introduction}

New particles with masses of $O(m_Z)$ or less must be weakly coupled
to electroweak gauge bosons and the electron vector and axial
currents in order to be in agreement with electroweak precision
data. Yet moderate couplings of these new states to other particles
are allowed, including to the Higgs boson, or to other new states
with masses of $O($TeV$)$. Production of the Higgs boson and new
states with TeV scale masses are expected to occur at the LHC and
their decays into the lighter particles can lead to a number of
striking signatures. Since the light states are weakly coupled, they
may be long-lived and have macroscopic decay lengths leading to
dramatic signatures of kinks in charged tracks \cite{scott} or
highly displaced vertices \cite{scott} \cite{strassler}. Either
phenomena can occur in models of low-energy gauge mediated
supersymmetry breaking \cite{scott}, and displaced vertices leading
to events with high multiplicity are predicted to occur in ``hidden
valley'' models \cite{strassler}. Direct decays of the Higgs boson
into these light states can significantly affect limits and search
strategies \cite{cascadehiggs}. Supersymmetry without $R$-parity can
also have striking signatures with displaced vertices occurring in
Higgs boson decay \cite{dekaplan}.

Virtual effects of the new physics at the TeV scale may also be
important, since the new physics presumably interacts (some what
strongly) with the Higgs boson, and probably top quarks. Integrating
out the new physics generates higher dimension operators involving
the Higgs boson, top quark, and any other light particles in the low
energy theory. Those operators that involve only the Standard Model
particles and the Higgs boson may significantly affect electroweak
LHC phenomenology and electroweak precision measurements
\cite{manoharwise}\cite{grojean} \cite{grinsteintrott}.

In \cite{mg} I described a scenario in which one important property
 of the Higgs boson - its lifetime -
  is significantly modified by the existence of
higher dimension operators at the TeV scale and additional light
states. In short, right-handed neutrinos are assumed to exist with
masses around the electroweak scale \footnote{For other, previous
literature on electroweak scale right-handed neutrinos, see
\cite{RHNEW1,RHNEW2,RHNHan,RHNEW4}. Higher dimension operators were
not considered by these authors.}. Active neutrinos are assumed to
acquire mass through a tree-level see-saw mechanism, which
necessitates tiny neutrino Yukawa couplings. The interesting
possibility of vanishing tree-level masses, which can occur for
certain neutrino couplings and nearly degenerate right-handed
neutrinos, is not explored here \cite{RHNEW1}. Under those
circumstances active neutrino masses are generated at one-loop,
allowing for larger neutrino couplings. Then the Higgs boson could
predominately decay into a right-handed and a left-handed neutrino
\cite{RHNEW1}. Here though the dominant decay mode of the Higgs
boson, and the mass scales of the right-handed neutrinos and their
decay channels are different from what I am exploring here.

Additional, unspecified physics at the TeV scale are assumed to
interact with the Higgs boson and the right-handed neutrinos. Then
below the TeV scale the new physics generates higher dimension
operators involving the Higgs boson and the right-handed neutrinos.
Due to these operators the Higgs boson can decay into right-handed
neutrinos if they are light enough. These decays will only dominate
over Standard Model decays if the Higgs boson mass is less than the
$WW$ mass threshold, and therefore only if at least one right-handed
neutrino is lighter than the $W$ gauge boson.

The dominant decay of the right-handed neutrino is into quarks and a
lepton. Since the right-handed neutrinos are long-lived, two
displaced vertices with average decay length any where from
$O(mm-10m)$ and larger occur, depending on the neutrino parameters
of the model. In the detector this would be visible as a highly
displaced vertex appearing to violate lepton number. The Higgs boson
decays into 6 particles, producing two such displaced vertices. In
this paper I discuss the collider bounds on such a scenario.
Attention is given to limits from the Tevatron, because their
kinematic reach is much larger than the LEP limit on a Standard
Model Higgs boson.

A diverse set of searches for new physics are considered, because
the nature of the new decay processes and the large Higgs boson
production rate suggest that several searches have the potential to
discover or exclude this scenario. Current Tevatron analyses are
already beginning to exclude regions of parameter space having short
average decay lengths $(\ltap O(2-5 cm))$ and $O(1)$ branching
fractions for the Higgs boson to decay into the right-handed
neutrinos. Decays of the Higgs into right-handed neutrinos having
such short decay lengths probably have to be subdominant. This might
occur naturally for some models, since such short decay lengths
require that the right-handed neutrino be heavier than approximately
$m_W$. If the Higgs boson is light it cannot decay into these
right-handed neutrinos, although it could decay into other, lighter
right-handed neutrino pairs which have longer average decay lengths.
Decay lengths greater than $O(10 cm)$ do not appear to be
constrained, yet their production rates are close to what can
currently be detected.

The outline of the paper is the following. Sections
\ref{sec:higgsdecay} and \ref{sec:RHNdecays} describes the
phenomenology of the decay of the Higgs boson and the right-handed
neutrinos, reviewing some results from \cite{mg} and introducing
notation. Model-independent branching fractions for decays of
right-handed neutrinos into inclusive final states are derived.
These are found to be quite useful in Section \ref{sec:tevatron},
which discusses possible Tevatron bounds on the new Higgs boson
decay processes.

The phenomenology of the Higgs boson decay just described occurs
when minimal flavor violation is used to determine the flavor
structure of the higher dimension operators \cite{mfv}. Since the
flavor properties of higher dimension operators involving only
right-handed neutrinos are weakly constrained, this assumption may
be overly restrictive for these particles. If it is relaxed, then
new Higgs boson decay channels are introduced, in which the Higgs
boson can decay into as many as 14 particles. Section
\ref{sec:beyondmfv} briefly discusses this drastically different
Higgs boson phenomenology.

Appendix \ref{sec:appA} discusses dimension 6 operators within the
context of minimal flavor violation and identifies new operators
that introduce a new but rare decay mode for the Higgs boson.
Appendix \ref{sec:appB} provides details of the right-handed
neutrino decays, with some attention given to final states having
quantum interference.

\section{Higgs Boson and Right-Handed Neutrino Decays}

\subsection{Higgs Boson Decays}
\label{sec:higgsdecay}

At dimension 5 the only phenomenologically relevant operator
involving right-handed neutrinos and the Higgs boson is \beq
\frac{c_{IJ}}{\Lambda} N_I N_J H^{\dagger} H ~. \label{o51} \eeq
Here $N_I$, $I=1,2,3$ are Majorana neutrinos (I will also refer to
them as ``right-handed neutrinos'') and $H$ is the Higgs boson with
vacuum expectation value (vev) $ \langle H \rangle =v/\sqrt{2}
\simeq 175 $ GeV. Other operators occur at dimension 5, but in the
context of minimal flavor violation they are naturally and
sufficiently suppressed \cite{mg}.

Interestingly, these operators (\ref{o51}) can be relevant for the
decay of the Higgs boson. For after electroweak symmetry breaking
the Higgs boson can decay \beq h \rightarrow N_I N_J \eeq with a
substantial rate. With $c_{IJ} \simeq O(1)$, these decays of the
Higgs boson into right-handed neutrinos can dominate over decays
into bottom quarks for a wide range of scales $\Lambda$.

Whether the coefficients $c_{IJ}$ of these operators are significant
depends on the physics at the scale $\Lambda$. Experimentally they
can be $O(1)$, since there are no constraints on these operators if
$\Lambda \simeq O($TeV$)$. One reason is that these operators
preserve custodial isospin and weak isospin, so there are no
constraints from electroweak precision data. In fact, after
electroweak symmetry breaking the only effect of these operators is
to shift the masses of the right-handed neutrinos. The tiny observed
left-handed neutrino masses do not constrain these operators either,
since contributions to the left-handed neutrino masses obtained from
inserting these operators into loops always appear with two powers
of the neutrino couplings, and are therefore always
subdominant\cite{mg}.

Yet it is well-known that generic dimension 5 and 6 operators at the
TeV scale involving only Standard Model fermions are excluded by
flavor-changing and $CP$ violating neutral current processes (FCNC),
and also by the tiny values of the left-handed neutrino masses. A
concern may be that any physics that suppresses such dangerous FCNC
operators to acceptable levels may simultaneously suppress the
operator (\ref{o51}). To investigate whether that occurs, in
\cite{mg} I applied the minimal flavor violation hypothesis
\cite{mfv} \cite{mfvquarks} \cite{mfvleptons} to the operators
(\ref{o51}). Under the assumptions of this hypothesis,  dimension 6
contributions to the $CP$ violating parameter $\epsilon$ can be
suppressed with $\Lambda \simeq 5-10$ TeV \cite{mfvquarks}, whereas
contributions to lepton number violating processes are sufficiently
suppressed for $\Lambda \simeq O($ TeV$)$ \cite{mfvleptons,mfvcg}.
Dimension 5 contributions to the left-handed neutrino masses are
also subdominant with $\Lambda \simeq O($TeV$)$ \cite{mg}.

``Minimal flavor violation" \cite{mfv} is a hypothesis about the
flavor structure of higher dimension operators and is a useful
framework to adopt here. With two key assumptions, the flavor
structure of higher dimension operators can be expressed in terms of
the quark, lepton and neutrino Yukawa couplings, as well as the
Majorana neutrino mass parameters. For the quarks and leptons, the
two assumptions are that the symmetry group is maximal, and that
there is only one order parameter for each flavor group. Relaxing
either hypothesis can lead to dangerous flavor violation, for then
the flavor structure of the higher dimension operators cannot be
completely specified by the Yukawa couplings. For instance, if there
is more than one spurion per symmetry group, then the linear
combination determining the Yukawa couplings is in general different
from the combination appearing in a higher dimension operator. Or,
if the symmetry group is not maximal then there are more group
invariants.

It is not surprising then that whether the operator coefficients
(\ref{o51}) are suppressed or not depends on the broken flavor
symmetry $G_N$ of the right-handed neutrinos \cite{mg}. In the case
of a maximal flavor symmetry, $G_N =SU(3)_N \times U(1)^{\prime}$
(see Appendix \ref{sec:appA} for definitions), the minimal flavor
violation hypothesis implies that \beq c_{IJ} =c \frac{
[m_R]_{IJ}}{\Lambda} + \cdots \eeq and is suppressed by the Majorana
neutrino mass matrix $m_R$. The reason for the suppression is that
both the Majorana masses and the operators (\ref{o51}) violate the
$SU(3)_R$ symmetry, that in minimal flavor violation is assumed to
be broken by only one order parameter.

If the broken flavor symmetry is smaller, then larger coefficients
are allowed. For instance if instead $G_N = SO(3)$ then \beq c_{IJ}
= c^{\prime} \delta_{IJ} + c^{\prime \prime} \frac{
[m_R]_{IJ}}{\Lambda} + \cdots \eeq where now $m_R$ is real. Here the
operator coefficients are not suppressed. This too isn't surprising,
since for universal couplings the operator (\ref{o51}) preserves an
$SO(3)$ symmetry, and is therefore allowed to be unsuppressed. Here
though the symmetry allows a large mass term for the right-handed
neutrinos of $O(\Lambda)$. To avoid such a heavy right-handed
neutrino I must make the technically natural assumption $m_R \ll
\Lambda$.

For either scenario, the couplings $c_{IJ}$ evaluated in the
right-handed neutrino mass basis are flavor-diagonal to all orders
in $m_R/\Lambda$. The decays $h \rightarrow N_I N_J$ are then flavor
diagonal $(I=J)$. But depending on the broken flavor symmetry of the
right-handed neutrinos, we see that the rate may either be
unsuppressed and universal, or suppressed and non-universal. These
distinctions affect both the branching fraction for the Higgs boson
to decay into these channels, and the lepton flavor dependence of
the final state. I refer the reader to \cite{mg} for additional
details.

One finds that with $c_{IJ}=\delta_{IJ}$,
\begin{eqnarray} \Gamma(h \rightarrow N_I N_I)
&=&\frac{v^2}{4 \pi \Lambda^2} m_h \beta^3_I
\end{eqnarray}
where $\beta_I$ is the velocity of $N_I$. If instead $c_{IJ}= M_I
\delta_{IJ}/\Lambda$, there is an additional suppression giving
\begin{eqnarray} \Gamma(h \rightarrow N_I N_I)
&=&\frac{v^2}{4 \pi \Lambda^2}\left(\frac{M_I}{\Lambda}\right)^2 m_h
\beta^3_I~.
\end{eqnarray}

In Figures \ref{fig:hdecaybb}(a) and \ref{fig:hdecaybb}(b) the ratio
${\cal R} \equiv \sum_I \Gamma [h \rightarrow N_I N_I ]/ \Gamma [h
\rightarrow b \overline{b}]$ is presented for $m_h=120$ GeV, a range
of $M_I$ and scale $\Lambda$. (The dip in the plots near $M_I \simeq
m_h/2 $ GeV is due to phase space suppression.) To simplify the
presentation of the plots, approximately universal right-handed
neutrino masses $M_I \simeq M_R$ are assumed. In Figure
\ref{fig:hdecaybb}(a) a universal coupling $c_{IJ} = \delta_{IJ}$ is
used, whereas in Figure \ref{fig:hdecaybb}(b) a suppressed coupling
$c_{IJ} = M_I \delta_{IJ}/\Lambda$ is assumed. Note that if the
coupling is unsuppressed there is a large range of parameter space
over which decays into right-handed neutrinos dominate decays to $b
\overline{b}$. This is because the decay to $b \overline{b}$ is
suppressed by the small bottom Yukawa coupling. In particular, the
decay to a right-handed neutrino dominates over the $b \overline{b}$
channel up to scales $\Lambda
>10 $ TeV (and up to $20$ TeV for $m_N \ll m_h$). (If Im$(c_{IJ})
\neq0$ the phase space for decays is larger \cite{mg}).  For
suppressed couplings, the ratios shown in Fig. \ref{fig:hdecaybb}(b)
indicate that decays into right-handed neutrinos are subdominant,
but not rare if the scale $\Lambda$ is low.

\FIGURE[t]{\centerline{a)\epsfxsize=3.0 in
\epsfbox{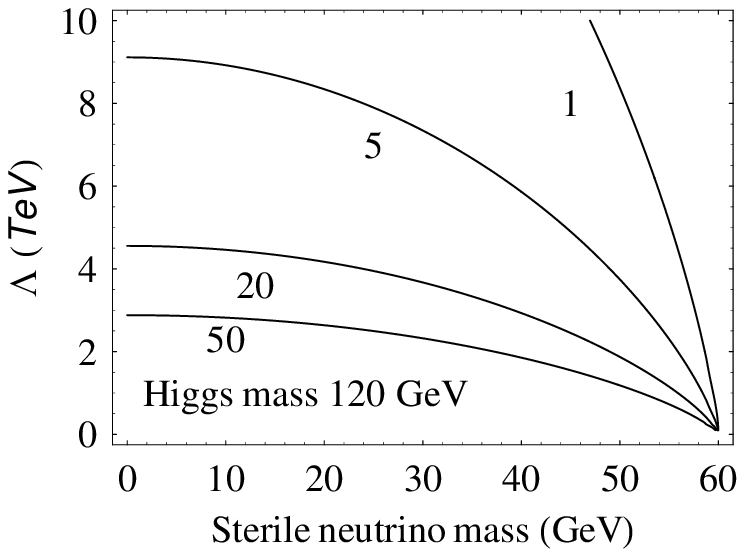} \hskip 0.5in b)\epsfxsize=3.0 in
\epsfbox{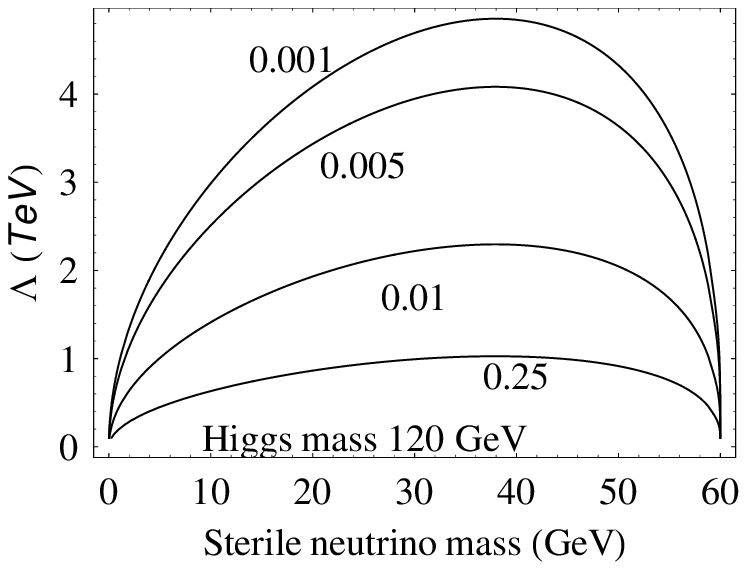}} \noindent \caption{Contour plots of
$R=\sum_I \Gamma[h \rightarrow N_I N_I]/\Gamma[h\rightarrow b
\overline{b}]$ for decay into 3 flavors of right-handed neutrino,
with $m_h=120$ GeV and in (a) $c_{IJ}=\delta_{IJ}$ and (b)
$c_{IJ}=M_I \delta_{IJ}/\Lambda$ . Note that in (b) the scale is
smaller. To simplify the presentation, right-handed neutrino masses
are assumed to be universal.}\label{fig:hdecaybb}}

If the Higgs boson mass is above both of the weak gauge boson mass
thresholds, then it can decay into $WW$ and $ZZ$ with large rates.
In this case decays of $h \rightarrow N_I N_I$ are subdominant, but
are certainly still interesting. In Figures \ref{fig:hdecayWW}(a)
and \ref{fig:hdecayWW}(b) the ratio ${\cal R} \equiv \sum_I \Gamma
[h \rightarrow N_I N_I ] /( \Gamma [h \rightarrow WW]+ \Gamma[ h
\rightarrow ZZ])$ is presented for $m_h=230$ GeV and a range of
right-handed neutrino masses and scales $\Lambda$. Again, for both
figures universal right-handed neutrino masses are assumed. In
Figure \ref{fig:hdecayWW}(a) a universal coupling $c_{IJ} =
\delta_{IJ}$ is used, whereas in Figure \ref{fig:hdecayWW}(b) a
suppressed coupling $c_{IJ} = M_I \delta_{IJ}/\Lambda$ is assumed.
Note that the rate for the Higgs boson to decay into all three
right-handed neutrino channels is not much smaller than for a decay
into the $WW$ and $ZZ$ channels. For example, with $\Lambda \simeq
2$ Tev and $O(1)$ coupling, the decay of the Higgs into three
right-handed neutrino flavors has a branching fraction of
approximately $20\%$, whereas for $\Lambda \simeq 10$ TeV it varies
from $0.01 \%-1 \%$ depending on $m_R$. For a suppressed coupling
and a low scale, $\Lambda \simeq 2$ TeV, the branching ratio is much
smaller, $O(10^{-4}-10^{-3})$. How much integrated luminosity is
needed to discover these decay processes at the LHC deserves further
study.

\FIGURE[t]{\centerline{a)\epsfxsize=3.0 in
\epsfbox{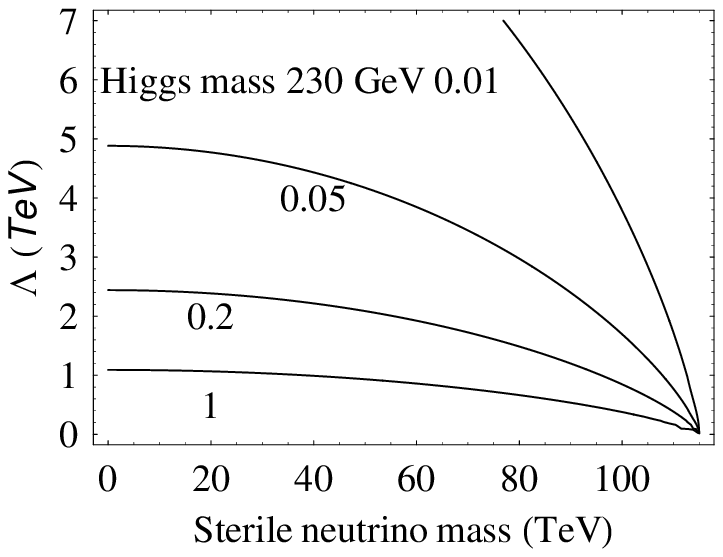} \hskip 0.5in b)\epsfxsize=3.0 in
\epsfbox{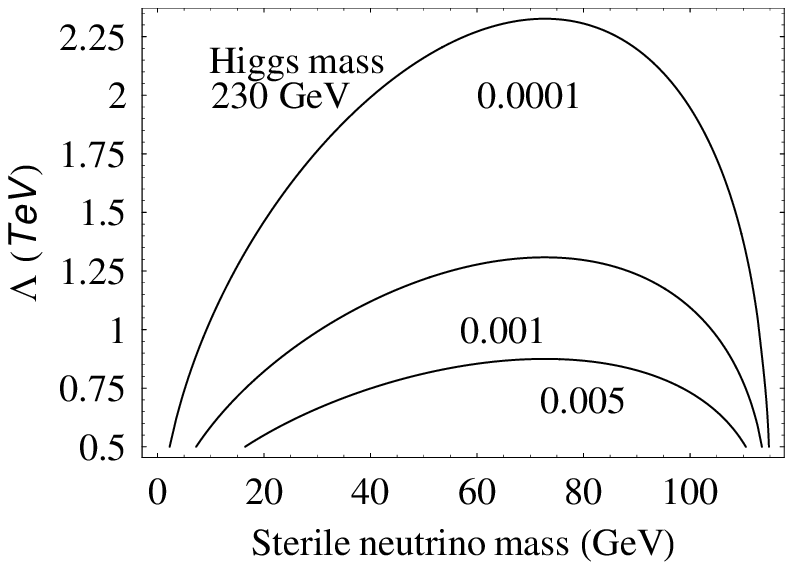}} \caption{Contour plots of $R=\sum_I
\Gamma[h \rightarrow N_I N_I]/(\Gamma[h \rightarrow WW]+ \Gamma[h
\rightarrow ZZ])$, with $m_h=230$ GeV. In (a) $c_{IJ} =\delta_{IJ}$
and in (b) $c_{IJ}= \delta_{IJ} M_I /\Lambda$. Note that in (b) the
scale is smaller. To simplify the presentation the right-handed
neutrino masses are assumed to be degenerate.}\label{fig:hdecayWW}}

\subsection{Right-Handed Neutrino Decays}
\label{sec:RHNdecays}
\subsubsection{Dominant Decays}

The standard see-saw mechanism introduces mass mixing between the
right-handed and left-handed neutrinos which leads to a mass matrix
for the active neutrinos given by
\begin{eqnarray}
&& m_L = \frac{1}{2}\lambda^T_{\nu} m^{-1}_R \lambda _{\nu} v^2 =
m_D ^T m^{-1}_R m_D ~.
\end{eqnarray}
Here $\lambda_{\nu}$ is the neutrino Yukawa coupling matrix, $m_D =
\lambda_{\nu} v /\sqrt{2}$ is the Dirac mass matrix, and $m_R$ is
the $3 \times 3$ right-handed neutrino mass matrix. We can choose a
basis where the latter matrix is diagonal and real with elements
$M_I$. In general they will be non-universal.

The active neutrino mass matrix is diagonalized by the PMNS matrix
$U_{PMNS}$ \cite{PMNS} to obtain the physical masses of the active
neutrinos, denoted by $m_I$. For generic Dirac and Majorana neutrino
masses no simple relation exists between the physical masses, the
rotations between the mass and gauge eigenstates, and the PMNS
matrix. An approximate relation for the neutrino couplings is \beq
 f_{I} \approx 7 \times 10^{-7} \left(\frac{m_I}{0.5
\hbox{eV}}\right)^{1/2} \left(\frac{M_I}{30 \hbox{GeV}}\right)^{1/2}
\label{approximateneutrinocoupling} \eeq where $\lambda_{\nu} = U_R
f U_L$ is expressed in terms of two unitary matrices $U_{L/R}$ that
diagonalize $\lambda^{\dagger} _{\nu} \lambda_{\nu}$ and
$\lambda_{\nu} \lambda^{\dagger}_{\nu}$,  and a diagonal matrix $f$
with elements $f_I$. In general $U_L$ is not the same as $U_{PMNS}$.

The mass mixing introduces couplings of the right-handed neutrinos
to the $W$ and $Z$, which at leading order in the Dirac masses are
obtained by replacing a left-handed neutrino $\nu_I$ with a
right-handed neutrino $N_J$ multiplied by the mixing matrix \beq
D_{IJ} =[m^T_D m^{-1}_R ]_{IJ}=[m^T_D]_{IJ} M^{-1}_J ~. \eeq  These
couplings are small, since they are suppressed by the tiny neutrino
couplings. If the right-handed neutrino masses are approximately
universal $M_I \simeq M$ and $U_R \simeq 1$, then
\begin{eqnarray} D_{JI} \simeq \sqrt{\frac{m_I}{M}}
[U_{PMNS}]_{IJ}=4 \times 10^{-6}\sqrt{\left(\frac{m_I}{0.5
\hbox{eV}}\right)\left( \frac{30 \hbox{GeV}}{M}\right)}
[U_{PMNS}]_{IJ} ~. \label{approximatemixingangles}
\end{eqnarray}

 Because of these couplings the right-handed neutrinos can
decay into Standard Model particles, albeit with a tiny rate. If the
right-handed neutrinos are heavy enough, the decays \beq N_I
\rightarrow W l_J, ~ Z \nu_J \label{Ndecay1} \eeq occur. If the
right-handed neutrinos are below the weak gauge boson mass
threshold, the following decays into three-body final states occur,
\beq N_I \rightarrow l_J W^* \rightarrow l_J f
\overline{f}^{\prime}~,~ N_I \rightarrow \nu_J Z^* \rightarrow \nu_J
f \overline{f} \label{Ndecay2} \eeq Since the right-handed neutrinos
are neutral and have Majorana masses, they can also decay into the
charge-conjugation of any of the final states in (\ref{Ndecay1}) and
(\ref{Ndecay2}).

Not surprisingly, the dominant decay is to final states containing
quarks. Model-independent branching fractions for inclusive decays
are discussed below and presented in Table
\ref{BFfordecaysintoquarks}. Decays into final states containing
charged leptons of specific flavor are of obvious experimental
interest, but those rates are model-dependent.

Since accurate branching ratios are important for either discovery
or setting limits, quantum interference effects should be included.
They occur in $N_I \rightarrow l_J \overline{l}_J \nu_J$ decays,
with the two leptons and the neutrino of the same flavor. This is
precisely the exclusive decay studied in the D0 search for displaced
muon pairs produced by the decay of a long-lived neutral particles
\cite{longlivedneutralD0}. Possible limits from their search is
discussed further in Section \ref{subsec:diplacedmuonpairs}. The
rate for decays $N_I \rightarrow l_J \overline{l}_J \nu_J$,
including the quantum interference, is discussed more in Appendix
\ref{sec:appB}.
%There are two final states with interfering diagrams that in both
%cases reduces the branching fractions. The rate for $N_I \rightarrow
%l_J \overline{l}_J \nu_J$ with the two leptons and the neutrino of
%the same flavor is reduced due to quantum interference between the
%charged current and neutral current exchange diagrams. This process
%is discussed more in Appendix \ref{sec:appB}.
In the contact limit $M_I \ll m_W$ the neutral current and charged
current diagrams destructively interfere, reducing the branching
ratio by $0.45$ compared to the incorrect result of adding the
diagrams incoherently.

Quantum exchange interference also occurs in final states with three
same-flavor neutrinos, $\nu_J \nu_J \overline{\nu}_J$, where the
branching ratio to this final state is reduced by approximately
$1/2$ in the contact limit. Here the interference is less important
since the rate for this final state is a fraction of the inclusive
decay rate $N_I \rightarrow$ nothing. Higgs exchange interferes with
neutral current exchange in decays $N_I \rightarrow b \overline{b}
+$missing energy. This process has not been included and is expected
to be small because of the tiny bottom Yukawa coupling.

\TABLE[h]{
\begin{tabular}{|c|c|c|}  \hline
final state &  & $F_{JKL}$ \\ \hline
  $l_J q \overline{q}^{\prime}$ & $ u \overline{d}  $ & $N_c c_W$
  \\
& $ c \overline{s} $ & $N_c c_W $ \\ \hline $\nu_J q \overline{q}$ &
$u \overline{u} + d \overline{d}+ s \overline{s}$ &
$\sum_i \left((g^{(i)}_L)^2+ (g^{(i)}_R)^2 \right)N^{(i)}_c c_Z= 1.5 c_Z$  \\
& $ c \overline{c}$ & $0.43 c_Z$ \\
  &  $ b \overline{b}$   & $0.55 c_Z$  \\ \hline
$l_J \overline{l}_{K \neq J} \nu_K $ &  & $c_W$    \\ \hline $\nu_J
l_J \overline{l}_J $ &  & $c_W + 0.13 c_Z +$interference term
$\rightarrow 0.59$  \\ \hline
 $\nu_J l_{K \neq J} \overline{l}_{K \neq J}$ &  & $
\left((g^{(K)}_L)^2+ (g^{(K)}_R)^2 \right)c_Z =0.13 c_Z$ \\ \hline
$\nu_J \nu_{K \neq J} \overline{\nu}_{K \neq J}$ & & $\frac{1}{4}
c_Z $ \\ \hline $\nu_J \nu_J \overline{\nu}_J $ & & $c_Z
(\frac{1}{4} + \hbox{quantum exchange interference}) \rightarrow
\frac{1}{8}$ \\ \hline total & & $N_{tot}(M_I)=8 c_W + 3.2 c_Z+0.72$
\\ \hline
\end{tabular}\caption{Right column gives the $F$ factor multiplying the
partial width (\ref{partialwidth}) for the final state listed in the
same row and no sum over $J$ or $K$. Quantum interference between
charged current and neutral current exchange occurs in final states
with two charged leptons and a neutrino of the same flavor (third
row from bottom). Final states with three same flavor neutrinos have
a quantum exchange interference term (bottom row). In both these
cases the interference is not insignificant below the gauge boson
threshold. For these two final states only the $F$ factor in the
contact interaction approximation is given (denoted by the long
arrows). See Appendix \ref{sec:appB} for descriptions of $c_W$,
$c_Z$ and more details of the computations of the partial rates.}
\label{Ffactor}}

One finds the partial decay rates
 \beq \Gamma[N_I \rightarrow f_J f_K  f_L  ]+ \Gamma[N_I
\rightarrow f^c_J f^c_K  f^c_L  ] = 2 \frac{G^2_F M^5_I}{192 \pi^3}
|D_{JI}|^2  F_{JKL} \label{partialwidth} \eeq Details are provided
in Appendix \ref{sec:appB}. The factor of $``2"$ occurs simply
because the right-handed neutrinos are Majorana particles, so at
tree-level they can decay with equal rates into a state and the
charge-conjugation of that state. The factor of $F_{JKL}$ for each
final state can be found in the last column of Table \ref{Ffactor}.
Using the results provided in Table \ref{Ffactor}, one obtains the
inclusive decay rate \beq \Gamma_{\hbox{total}}[N_I]= 2 \frac{G_F^2
M^3_I}{192 \pi^3} [m_D m^{\dagger}_D]_{II} N_{tot}
\label{Ntotalrate} \eeq where $N_{tot}(M_I)=8 c_W +3.2 c_Z +0.72$.
 The
functions of $c_W$ and $c_Z$ are described in Appendix
\ref{sec:appB} and depend on the right-handed neutrino mass $N_I$.
They are given by (\ref{cG}) and describe the effects of finite
momentum transfer. (The term $``0.72"$ in $N_{tot}$ represents the
contributions from same-flavor charged lepton $l_J \overline{l}_J
\nu_J$ and same-flavor neutrino $\nu_J \nu_J \overline{\nu}_J$ final
states to the total rate. Unlike all the other final states, each of
these final states have quantum interfering contributions which have
been computed in the low-energy limit $M_I \ll m_W$; more details
can be found in Appendix \ref{sec:appB}.)

Let me now comment on some of the terms appearing in the inclusive
decay rate. After summing over all the final states the dependence
of each partial rate on the couplings $D_{JI}$ can be expressed in
terms of the Dirac mass matrix $m_D$ and right-handed neutrino
masses $M_I$. This changes the overall dependence of the rate from
$\Gamma \propto M^5_I$ to $\Gamma \propto M^3_I$. Additional
dependence on the right-handed neutrino mass occurs through the
implicit dependence of the Dirac mass on the left-handed and
right-handed neutrino masses.

\TABLE[h]{\begin{tabular}{|c|c|} \hline final state  & Branching Fraction \\
\hline light quark flavors + charged lepton & $6 c_W/N_{tot} \simeq
0.50$
\\ \hline light quarks + missing energy & $1.5 c_Z/N_{tot} \simeq 0.13$
\\ \hline $c \overline{c}$ + missing energy &
$0.43 c_Z/N_{tot} \simeq 0.036$ \\
\hline $b \overline{b}$ + missing energy & $0.55 c_Z/N_{tot} \simeq
0.046$
\\ \hline two charged leptons and missing energy
& $(2 c_W + 0.59+0.26 c_Z)/N_{tot} \simeq 0.24$ \\
\hline neutrinos
& $\approx(1/8+ c_Z/2)/N_{tot} \simeq 0.05$ \\
\hline
\end{tabular}\caption{Inclusive branching fractions for decays of
$N_I$ into final states containing quarks, dileptons plus missing
energy, or nothing. ``Charged leptons'' refers to summing over $e$,
$\mu$ and $\tau$. These branching fractions are independent of the
Dirac mass matrix elements. Processes having quantum interference
were obtained in the low-energy limit $M_I \ll m_W$. Numbers
displayed to the right-side of the right column are obtained in the
same limit $M_I \ll m_W$, where the $c's \simeq 1$.
}\label{BFfordecaysintoquarks}}

Inspecting Table \ref{Ffactor} and the total decay rate
(\ref{Ntotalrate}) one notices that the branching fractions for some
inclusive processes are independent of the Dirac mass matrix
elements, since the factor of $[m_D m^{\dagger}_D]_{II}$ cancels in
the ratio. In particular, model-independent branching fractions can
be obtained for inclusive decays into quarks with any charged lepton
or with missing energy, or decays into nothing. These branching
fractions are listed in Table \ref{BFfordecaysintoquarks}. The
dominant decay mode is semi-leptonic, with the right-handed neutrino
decaying into light quark flavors and a charged lepton of any
flavor, including $\tau$. It occurs with a branching fraction of
roughly $1/2$. The next dominant hadronic decay mode is purely
hadronic. Here the right-handed neutrino decays into light quarks
and missing energy with a branching fraction of approximately 0.13.
The right-handed neutrino does not have an exclusive decay into $b
\overline{b}$. Instead the decay is $N_I \rightarrow b
\overline{b}+$missing energy and occurs with a branching fraction of
approximately 0.05. The decay $N_I \rightarrow$nothing also occurs
with a branching fraction of roughly $0.05$. Finally, the inclusive
decay of a right-handed neutrino into missing energy and two charged
leptons of opposite-sign but unspecified flavor has a branching
fraction of approximately $0.24$.

The rates for decays into final states containing charged leptons of
specific flavors {\em is} model-dependent. For decays into events
containing charged leptons of specific flavor $J$ depend on the
Dirac mass matrix element $|[m_D]_{JI}|^2$, which in general cannot
be expressed in terms of the active neutrino masses, the
right-handed neutrino masses and the PMNS matrix. This feature may
be viewed as an experimental opportunity, for a measurement of these
branching fractions directly determines these matrix elements, which
cannot be obtained from any low-energy experiment involving the
active neutrinos.

Converting the decay rate to an average lifetime gives
 \beqa c \tau _I &=& 0.95 m \left(\frac{1.49}{c_W + 0.40
c_Z+0.09}\right) \left(\frac{\hbox{30 GeV}}{M_I}
 \right)^3 \left(\frac{(120 \hbox{ keV})^2}{[ m_D m^{\dagger}_D]_{II}}
 \right) \label{decaylengthexact}
 \eeqa
This is an exact relation as no assumptions about the right-handed
neutrinos masses or the Dirac matrix have been made. As already
noted, diagrams having quantum interference are approximated by
their low-energy values $M \ll m_W$ and represent the $``0.09"$
contribution appearing in the formulae above \footnote{As noted in
\cite{mg}, there final states having quantum interference were
simply added incoherently. This difference accounts for the $\approx
5\%$ discrepancy between (\ref{decaylength1}) and the equivalent
formula given in \cite{mg}.}. The lifetime is seen to be very
sensitive to the right-handed neutrino mass : changing the
right-handed neutrino mass by a factor of 2 changes the average
lifetime by over an order of magnitude. Longer lifetimes occur for
smaller neutrino couplings, which are correlated with the active
neutrino masses.

To relate the lifetime to a measured neutrino mass requires a
specific model, since the factor $[m_D m^{\dagger}_D]_{II}= [U_R f^2
U^{\dagger}_R]_{II}$ is in general not simply related to the active
neutrino masses. A simple-minded approximation is $[m_D
m^{\dagger}_D]_{II} \approx  m_I M_I$ which gives  \beqa
 c \tau_I & \approx & 1 m \left(\frac{1.49}{c_W + 0.40
c_Z+0.09}\right) \left(\frac{\hbox{30 GeV}}{M}
 \right)^4 \left(\frac{0.5 \hbox{eV}}{m_{I}} \right)
 \label{decaylengthapproximated} \eeqa
Using (\ref{decaylengthapproximated}), Figure \ref{decaylength1}
shows a range of active neutrino masses $m_I$ and right-handed
neutrino masses giving a variety of lifetimes.

From Figure \ref{decaylength1} it is seen that parameters having a
light Higgs boson $($i.e, $m_N < m_h/2<80$GeV$)$ and $c
\tau=O(5cm-4m)$ correspond to active neutrino masses of
$O(0.05-0.5)$eV. For most of this range the active neutrinos masses
must be degenerate in order to be consistent with what is known
about neutrino mass differences. This is an interesting scale, for
future experiments may be sensitive to some of this region. The
lower end of this range is consistent with either a normal or
inverted hierarchy, with the absolute scale approximately given by
the atmospheric neutrino anomaly, $\Delta m^2_{32}=(1.9-3.0) \times
10^{-3} \hbox{eV}^2 \simeq (0.05)^2 \hbox{eV}^2$ \cite{PDG}.

The average decay length of a right-handed neutrino depends on its
lifetime and its velocity. In the rest frame of the Higgs boson,
$\gamma \beta$ for the right-handed neutrino ranges from $0.7-5$ for
$M_I/m_h$ ranging from $0.4-0.1$. Average decay lengths are then not
significantly different from $O(c \tau)$ unless the right-handed
neutrinos are much lighter than the Higgs boson.

 I end this section
by reiterating that relations between the active neutrino masses and
the average decay lengths in a specific model will differ from the
comments of the last paragraph and curves shown in Figure
\ref{decaylength1}. The reason is that Figure \ref{decaylength1}
uses (\ref{decaylengthapproximated}) which is a simple-minded
relation between the Dirac matrix elements and the active neutrino
mass. The model-independent result is given by
(\ref{decaylengthexact}) and depends only on $[m_D
m^{\dagger}_D]_{II}$. While this matrix element is correlated with
the active neutrino masses, the relationship is more complicated in
general.

\FIGURE[t]{\epsfxsize=4.0 in \epsfbox{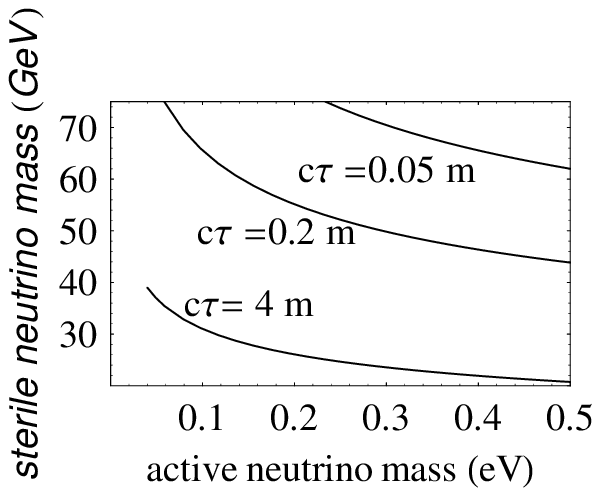}
\caption{Average lifetime from (\ref{decaylengthapproximated})
assuming a simple-minded relation between the Dirac matrix elements
and active neutrino mass. The contours are cropped at $m_I = 0.05$
eV. For smaller values of the active neutrino mass, the right-handed
neutrino mass needed to obtain $c \tau=O(0.1m-4m)$ is off the scale
and not shown for convenience.}\label{decaylength1}}

\subsubsection{Subdominant Decays}
\label{subdominantdecays}

It was seen that a dimension 5 operator can significantly modify the
phenomenology of the Higgs boson. Could other higher dimension
operators or radiative corrections cause subdominant decays that
nonetheless might be important to consider? In \cite{mg} it was
argued that there are no other important dimension 5 operators. At
dimension 6 there are a large number of operators involving
right-handed neutrinos and Standard Model particles. These may be
found in Appendix \ref{sec:appA}, where the size of their
coefficients are estimated using minimal flavor violation. Although
there are a number of operators, most of them are irrelevant either
because they are suppressed by two Yukawa couplings, or because they
contribute at a subdominant rate to the processes (\ref{Ndecay1}) or
(\ref{Ndecay2}) previously discussed.

There is however a new process that occurs from two dimension 6
operators and a 1--loop process. After electroweak symmetry breaking
there are two one--loop diagrams that together generate a magnetic
and an electric moment operator for a right-handed neutrino and an
active neutrino. Similarly,  two of the dimension 6 operators
generate the same operator, but suppressed by the scale $\Lambda$.
Together they combine into the following magnetic moment operator
\beq
 e c_M \frac{1}{16 \pi^2 v^2}
 (N m_D \sigma^{\mu \nu} \nu_L) F_{\mu \nu}
  \label{magneticmomentop}
\eeq where $F_{\mu \nu}$ is the electromagnetic field strength,
$\sigma^{\mu \nu} =\sigma_2(\sigma^{\mu}\overline{\sigma}^{\nu} -
(\mu \leftrightarrow \nu))/4$. The coefficient of the operator is a
sum of two contributions, one from a one-loop process and the other
from two dimension 6 operators :\beq c_M = c_{1-loop} + c^{(6)}_M
\frac{16 \pi^2 v^2}{\Lambda^2} ~. \eeq The one-loop contribution
$c_{1-loop}$ is calculable and expected to be $O(1)$. The
coefficient $c^{(6)}$ is a linear combination of the coefficients of
the dimension 6 operators described in Appendix \ref{sec:appA}. The
size and flavor structure of the dimension 6 operators has been
estimated using minimal flavor violation. With $c_{1-loop}$ and
$c^{(6)}_M$ of the same size, $c_M$ is dominated by the contribution
from the higher dimension operators for $\Lambda \ltap 4 \pi v = 3$
TeV.

This operator causes the decay \beq N \rightarrow \gamma
\nu_L~,~\gamma \overline{\nu}_L ~. \eeq Note that these processes
are not parametrically suppressed by Yukawa couplings compared to
(\ref{Ndecay1}) and (\ref{Ndecay2}), since all of these amplitudes
are $O(m_D)$. Summing over the three light left-handed neutrinos and
anti-neutrinos, one finds the inclusive rate \beq \Gamma[N_I
\rightarrow \gamma +\hbox{missing}] = \frac{\alpha_{em}}{2} | c_M|^2
 \frac{[m_D m^{\dagger}_D]_{II}}{(4 \pi v)^4} M_I^3 ~. \eeq
This rate has the same parametric dependence on the Dirac matrix
elements and the right-handed neutrino mass as the rate for the
dominant decay process (\ref{Ntotalrate}).
 Comparing
the two gives the following inclusive branching fraction, \beq
Br(N_I \rightarrow \gamma +\hbox{missing}) = \alpha_{em} | c_M|^2
\frac{192 \pi^3}{16} \frac{1}{(16 \pi^2)^2} (c_W + 0.4
c_Z+0.09)^{-1} \simeq 7\times 10^{-5}  |c_M|^2 \label{Nnugammadecay}
\eeq Like the inclusive decays given in Table
\ref{BFfordecaysintoquarks}, this branching fraction is independent
of the Dirac mass matrix elements, the right-handed neutrino's mass
and its flavor. Although this branching fraction may seem tiny, it
is much larger at small $\Lambda$. In particular, for $\Lambda \ltap
3$ TeV and $c^{(6)}_M \simeq c_{1-loop}$, \beq Br(N_I \rightarrow
\gamma +\hbox{missing}) \simeq 0.007 |c^{(6)}_M|^2 \left(
\frac{\hbox{TeV}}{\Lambda} \right)^4 \eeq which is sizable.

 Constraints on the operator (\ref{magneticmomentop}) occur from
Tevatron data, which are discussed in Section
\ref{sec:longlivedneutrals}. The CDF search for delayed photons
\cite{longliveddecaytophotons} is beginning to exclude
$c^{(6)}_M=1$, $\Lambda \ltap 1.3 ~\hbox{TeV}$ if the dominant decay
mode of the Higgs boson is to {\em non-relativistic} right-handed
neutrinos. Because of the sensitivity of the branching fraction to
the scale $\Lambda$, $\Lambda \gtap 2$ TeV is allowed. Constraints
from low-energy experiments are weaker. For instance, although at
tree-level this operator contributes to the left-handed neutrino
magnetic moment because of mass-mixing between the left and
right-handed neutrinos, it is tiny. One finds a neutrino magnetic
moment of the size $\mu _{\nu_L} \simeq m_L m_e \mu_B /\Lambda^2
~\ltap 10^{-18} \mu_B (\hbox{TeV}/\Lambda)^2 $ for $\Lambda \ltap 4
\pi v$, which is well-below the experimental limit (recall that in
my notation $m_L$ is the active neutrino mass matrix, not the
charged lepton masses.) .

\subsection{Back to Higgs Decays: Summary}
\label{sec:higgssummary}

 Higgs bosons can decay to right-handed
neutrinos with a branching fraction that depends on the Higgs boson
mass, the coefficient of the operators (\ref{o51}) and the scale
$\Lambda$. In the favorable situation with unsuppressed couplings
and $m_h <2 m_W$, these decays can dominate over $h \rightarrow b
\overline{b}$. Specific branching ratios are shown in Figures
\ref{fig:hdecaybb} and \ref{fig:hdecayWW} for a number of scenarios.
The right-handed neutrinos are long-lived due to their tiny
couplings with Standard Model particles. They have displaced
vertices (\ref{decaylengthexact}) whose values depend on the
neutrino mass parameters, and typically range from $O($mm$)$ to
$O(10$m$)$ or larger. The average decay length is extremely
sensitive to the mass of the right-handed neutrinos
\ref{decaylengthexact}, with heavier right-handed neutrinos having
shorter decay lengths. The right-handed neutrinos decay
predominantly into quarks and a charged lepton or missing energy.
Branching fractions for some inclusive processes are given in Table
\ref{BFfordecaysintoquarks}. The right-handed neutrinos may also
decay into a photon and neutrino, with an inclusive branching
fraction given by (\ref{Nnugammadecay}). There is a calculable
contribution to this decay rate which, barring accidental
cancelations between it and contributions from higher dimension
operators, sets a lower bound to the branching fraction for this
decay.

\section{Experimental Limits: Tevatron}
\label{sec:tevatron}

 Since the left-right mixing angles are
extremely tiny, of $O(10^{-6}-10^{-5})$ and only right-handed
neutrino masses $M_I>O($GeV$)$ are considered, constraints from
neutrinoless double-$\beta$ decay, measurements at the $Z$-pole, or
from Drell-Yan production of right-handed neutrino \cite{RHNHan}, do
not exist. Cosmological constraints do not exist either, since with
$M_I \gtap O($GeV$)$ the right-handed neutrinos decay before
Big-Bang-Nucleosynthesis.

Right-handed neutrinos are most effectively produced from the decay
of a Higgs boson (see Section (\ref{sec:higgssummary}) for a brief
summary). As discussed in the previous section, the dominant decay
of the Higgs boson may be into two long-lived right-handed
neutrinos, with average decay lengths $\gamma \beta c \tau \simeq
O(mm-10m)$ (or larger) in an interesting range. These decay lengths
depend on the right-handed neutrino masses and the Dirac masses.
Decays occur inside the detector for a wide range of parameters, but
each right-handed neutrino will in general have a different average
decay length.

Whether these decays dominate depends on the sizes of the
coefficients of the operator (\ref{o51}). Each right-handed neutrino
decays into three particles, some of which may be charged. Its
dominant decay is into a charged lepton and a pair of quarks (see
Table \ref{BFfordecaysintoquarks}). The main signature for these
events is a displaced quark pair and a charged lepton which
reconstruct to a secondary vertex. Subdominant decays can produce
two charged leptons at each secondary vertex. Further, each Higgs
decay has two such secondary vertices. Since most of the decays are
into visible energy, and the mass of the Higgs boson is spread
across six final state particles, the event is additionally
characterized by little missing energy and low $p_T$ for each
particle. Events with higher $p_T$ and missing energy occur when the
Higgs boson is produced in association with a $W$ or $Z$.

The Tevatron currently does not have any direct searches for this
type of signal. A direct search for this signal will probably
require several complementary searches, since a wide range of
average decay lengths is expected and no one search can cover all
possible values. Moreover, in a given model the three right-hand
neutrinos will in general have different average decay lengths, so a
single search may only be sensitive to part of the spectrum.

There are a number of existing searches for new physics that have
similar, although not identical, signatures. These are the searches
for $R-$parity violation in supersymmetric models, searches for
long-lived neutral particles, inclusive searches for events
containing same-signed dileptons and searches for anomalous events
containing photons. These searches are beginning to place some
bounds on the parameters of the model discussed here. Given that,
more accurate analyzes of the processes described below are needed.
It is clear that interesting bounds could be obtained with more
integrated luminosity combined with a search optimized for the
signal.

\subsection{$R_p$-violating Searches}

Since the Higgs boson can decay into 3 and 4 leptons with little
missing energy, searches for $R$-parity violation from the $``LLE"$
operator may have a decent efficiency for selecting these events. In
particular, the CDF \cite{CDFRp} and D$0$ \cite{D0rp} searches for
this type of R-parity violation look for 3 charged leptons of the
form $lll$ or $ll \tau$ where $l=e$ or $\mu$ and there is no
requirement on the flavors of $l$. Importantly, there is no cut on
missing energy.

In the $R$-parity violating model the leptons are produced from the
decay of the lightest neutralino through the gaugino-lepton-slepton
and R-parity violating $``LLE"$ operators. To set limits, two
minimal supergravity models were considered where the parameters
were chosen such that chargino-pair and chargino-neutralino
production, rather than cascade decays of squarks and gluinos to
neutralinos and charginos, were the main source of leptons. For
operators involving only the first and second generation the limits
on the production cross-section to these final states are quite
strong, of the order $ 0.05$ pb for D0, $\simeq 0.07-0.1$ pb for
CDF, and are approximately independent of the mass of the lightest
neutralino.
%Limits involving one third generation are about an order
%of magnitude weaker.

In direct production of the Higgs boson, charged leptons are
produced in the cascade decay $h \rightarrow N_I N_I \rightarrow l
+\cdots$. Since the decay of a single right-handed neutrino can
produce at most two charged leptons, four charged leptons only occur
if both right-handed neutrinos decay this way. Three charged leptons
are produced if one right-handed neutrino decays to a single charged
lepton and the other decays to two. From Table
\ref{BFfordecaysintoquarks}, the branching fraction for a
right-handed neutrino to decay inclusively into jets and a charged
lepton, summed over all lepton flavors including $\tau$, is
approximately 0.5. The branching fraction for it to decay into two
charged leptons and missing energy, summed over all flavors, is
approximately 0.2. The branching fraction for the two right-handed
neutrinos to decay into three or more charged leptons, summed over
all flavors, is then $2(1/2)(0.2)+(0.2)^2=0.24$.

Another source of charged leptons is if the Higgs boson is produced
in association with a $W$ or $Z$ boson that decays leptonically. If
$W \rightarrow l \nu$ occurs, then the two right-handed neutrino are
required to produce at least two charged leptons, rather than three.
If $Z \rightarrow l \overline{l}$, then the two right-handed
neutrinos need only produce at least one charged lepton. One finds
the branching fraction for two right-handed neutrinos to decay into
at least two (one) charged leptons, summed over lepton flavors, is
approximately $0.6 (0.9)$.

In sum the rate for producing three or more charged leptons is \beqa
\sum_{\hbox{lepton flavor}} \sigma(p+\overline{p} \rightarrow h +Y
\rightarrow l^{\prime} l^{\prime \prime} l^{\prime \prime \prime}
+X) &\simeq&  \left[2.5 (0.24) \hbox{pb} + (0.3 \hbox{pb})(0.2)(0.6)
\right. \nonumber \\ & & \left.+(0.2
\hbox{pb})(0.06)(0.9) \right] \sum_I Br(h \rightarrow N_I N_I) \nonumber \\
& \leq & 0.63 \hbox{pb} \label{rp1} \eeqa This result includes
decays to $\tau$ leptons, which were needed to obtain a
model-independent result. Since $l \tau \tau$ and $ \tau \tau \tau$
final states were not included in either searches, this result
overestimates the number of events eventually accepted by the
searches. It should be interpreted as an upper bound that is
independent of the neutrino mass parameters. I have also included in
the first term events from associated Higgs production where the
gauge bosons do not decay to charged leptons. Higgs production
cross-sections are estimated as $2$ pb for gluon-gluon fusion, $0.3$
pb for associated $W$ production, and $0.2$ pb for associated $Z$
production \cite{higgsproduction}. The reader will notice that the
estimate above is about an order of magnitude larger than the limits
obtained for the R$-$parity violating models.

I will use these values to set limits on decays of the Higgs boson
into right-handed neutrinos. To do this I will assume that the
acceptances for the Higgs boson and R-parity violating decays are
similar. In actuality, there are two factors which reduce the
acceptance. Most of the signal events are from gluon-gluon fusion
Higgs production. But since the Higgs boson decay products are soft
(6 body decay), the acceptance for these events will be lower than
in the 3-body neutralino decay into leptons (in the D0 search, the
lightest neutralino mass is varied over $\approx 90-140$ GeV, which
is comparable to the mass range for the Higgs boson considered
here). A weaker limit is then expected, but a detailed study is
needed. Events from associated production of the Higgs boson are not
expected to have significantly different acceptance, since there is
one charged lepton (or 2 if from $Z$ decay) with high $p_T$.
However, here the net cross section is smaller, about $0.03$ pb
which is below the limit set on the $R-$parity violating models.

The other condition reducing the acceptance is the following. Both
experiments require that a number of the leptons be prompt. This
means that the minimum separation between the lepton and the primary
vertex projected onto the transverse plane must not exceed some
value $d_0$. Specifically, for this analysis CDF requires that $d_0
<0.2$ cm and $D0$ requires that $d_0 < 2$ cm. A right-handed
neutrino with average decay length $D_I \gg O(d_0)$ will have a much
lower acceptance, since most of the decays will have a minimum
transverse distance exceeding $d_0$. However, since the decay occurs
at random, the location of the secondary vertex forms a distribution
and a small fraction of events will decay with a transverse distance
$d_T <d_0$. To estimate that, I will crudely approximate the allowed
region as a sphere of radius $R < O(d_0)$. This is a crude
approximation, since actual detectors are tubular rather than
spherical, and moreover the requirement is only on the transverse
distance, not the physical distance. Then the probability that a
right-handed neutrino with average decay length $D_I$ decays within
a sphere of size $R=d_0$ is approximately \beq P(d<d_0;D_I) = 1-
e^{-d_0/D_I} ~.\eeq The average decay length $D_I$ depends on the
right-handed neutrino flavor and on its velocity $\beta_I$ and boost
$\gamma_I$. It is given by \beq D_I = \gamma_I \beta_I c \tau_I
~.\eeq

The estimate (\ref{rp1}) for the inclusive production cross-section
for three or more charged leptons is then modified to \beqa
\sum_{\hbox{lepton flavor}} \sigma(p+\overline{p} \rightarrow h +Y
\rightarrow l^{\prime} l^{\prime \prime} l^{\prime \prime \prime}
+X) & \simeq & \sum_I Br(h \rightarrow N_I N_I) \left[0.6
P(d<d_0;D_I)^2 \right.
\nonumber \\
& & \left. + 0.04 P(d<d_0;D_I)  \right] \hbox{pb} \eeqa This result
is maximized when all of the right-handed neutrinos have the same
average decay length $D_I = D_*$ (see the discussion below
(\ref{dilepton-total})). Then \beqa \sum_{\hbox{lepton flavor}}
\sigma(p+\overline{p} \rightarrow h +Y \rightarrow l^{\prime}
l^{\prime \prime} l^{\prime \prime \prime} +X) & \leq & \left[0.6
P(d<d_0;D_*)^2 \right.
\nonumber \\
& & \left. + 0.04 P(d<d_0;D_*)  \right] \hbox{pb} \nonumber
\\ & & \times \sum_I Br(h
\rightarrow N_I N_I) \label{rp-total} \eeqa

The stronger limit is from D0, since they have a weaker cut on
$d_0$. Assuming the signal here has the same acceptance as the
signal in the R$-$parity violating models, then with $D_* \gtap 10$
cm and $\sum_I Br(h \rightarrow N_I N_I)=1$, the production cross
section is $ \leq 0.03$ pb which is less than but close to the $D0$
bound. The actual limit may be weaker than this estimate for several
reasons. The acceptance for events from primary Higgs production may
be lower, since the leptons produced in the Higgs boson decay are
softer than those produced in the decay of the lightest neutralino.
Events from associated Higgs production may not have a lower
acceptance, but their rate, given by the second term in
(\ref{rp-total}), is much smaller. And as mentioned above, this
result overestimates the signal since it includes all the decays of
right-handed neutrinos to $\tau$s.

In a specific model the production cross section may not saturate
the model-independent upper bound derived above. For instance, if
the average decay lengths are hierarchical then only one
right-handed neutrino may have a significant contribution to
(\ref{rp-total}). If so, the total rate is suppressed by the
branching for the Higgs boson to decay into those right-handed
neutrinos.

\subsection{Inclusive Di-lepton Searches}

CDF has an inclusive search for di-lepton events where the two
leptons have the same charge \cite{CDFsamesignlepton}. Since the
decay of right-handed neutrinos can produce leptons of the same
sign, in principle the branching fraction for decays of the Higgs
boson into right-handed neutrinos could be constrained by this
search. These events occur from a number of processes and a
model-independent upper bound on their production rate is derived
below.

First consider primary production of the Higgs boson which then
decays to right-handed neutrinos. A single right-handed neutrino can
produce a single charged lepton and quarks with a high branching
fraction. With a smaller branching fraction it can decay to two
charged leptons and missing energy. Here though the two charged
leptons have the opposite sign. Events with two charged leptons of
the same-sign therefore requires that both right-handed neutrinos
decay leptonically. From Table \ref{BFfordecaysintoquarks}, the
branching fraction for a right-handed neutrino to decay inclusively
into jets and a charged lepton, summed over all charged lepton
flavors (including $\tau$), is approximately 0.5. The probability
for the two right-handed neutrinos to decay into jets and two
charged leptons of the same sign and summed over all 3 lepton
flavors is therefore $(0.5)^2/2=1/8$. The branching fraction for a
single right-handed neutrino to decay inclusively into two charged
leptons, summed over charged lepton flavors, is approximately 0.21,
so the branching fraction for two right-handed neutrinos to give
three charged leptons, summed over all flavors, is
$(0.5)(2)(0.21)=0.21$. In this case two of the leptons will have the
same charge. Finally, the branching fraction to obtain four charged
leptons and missing energy, summed over all charged lepton flavors,
is approximately $(0.2)^2=0.04$. In total, the branching fraction
for two right-handed neutrinos to decay into at least two charged
leptons of the same sign and summed over all lepton flavors is
approximately $0.37$. The rate for producing two {\em prompt}
same-signed di-leptons from primary Higgs production and summed over
all lepton flavors is then \beqa  \sum_{\hbox{lepton flavor}}
\sigma(p + \overline{p} \rightarrow
 h +Y \rightarrow  l^{\pm} l^{\prime \pm}+X)_{prmy} & = & (2.5
\hbox{pb})(0.37) \nonumber \\
& & \times \sum_I Br(h \rightarrow N_I N_I) (P(d<d_0;D_I))^2 \nonumber \\
&=& 0.93 \hbox{pb} \nonumber \\
& & \times  \sum_I Br(h \rightarrow N_I N_I) (P(d<d_0;D_I))^2
\nonumber \\ & & \label{dilepton-primaryHiggs} \eeqa The two factors
of $P(d<d_0,D_I)$ estimates the probability that both of the
right-handed neutrinos, each of average decay length $D_I$, decay
inside the prompt signal region $d_0 < 2$cm. By using a Higgs boson
production cross-section of $\approx 2.5$ pb for $m_h \simeq 115$
GeV I am also including associated Higgs boson production with an
electroweak gauge boson (which contributes in total $\approx 0.5$ pb
to the Higgs boson production cross-section).

Events with same-signed leptons also occur with associated Higgs
boson production, where one of the gauge bosons decays leptonically.
The efficiency for events from associated production will be higher,
since they have an energetic charged lepton, and if the lepton is
from a $W$, significant missing energy. If  $W \rightarrow l^{\pm}
\nu (\overline{\nu})$ occurs, then only one right-handed neutrino is
required to decay into a charged lepton of the same sign. If $Z
\rightarrow l^- l^+$, the right-handed neutrino can decay to a
charged lepton of either sign, although the rate for these events is
small. The branching fractions for a right-handed neutrino to decay
into a single charged lepton and summed over all lepton flavors is
approximately 0.5, and to decay into two charged leptons and summed
over all lepton flavors is 0.21. Next, the probability that at least
one of the two right-handed neutrinos has a prompt decay is
$2P(d<d_0;D_I)(1-P(d<d_0;D_I))$.(Events where both right-handed
neutrinos decay promptly was included in the previous result
(\ref{dilepton-primaryHiggs}). To include it here would be to double
count.) Then the contribution from associated production, where the
electoweak gauge boson decays leptonically and only one right-handed
neutrino decays promptly, is given by \beqa \sum_{\hbox{lepton
flavor}} \sigma(p+ \overline{p} \rightarrow h+W/Z \rightarrow
l^{\pm} l^{\prime
 \pm}+X) &= & \left[\frac{1}{2}(0.2)(0.3 \hbox{pb})(0.5)+
(0.2)(0.3 \hbox{pb})(0.21)+ \right. \nonumber \\ & & \left. +
(0.06)(0.2 \hbox{pb})(0.72) \right] \nonumber \\
& & \times \sum _I Br(h \rightarrow N_I N_I) \nonumber \\ & & \times
2(P(d<d_0;D_I))(1-P(d<d_0;D_I)) \nonumber
\\ &=& 0.072 \hbox{pb} \nonumber \\
& & \times  \sum _I Br(h \rightarrow N_I N_I)
P(d<d_0;D_I) \nonumber \\ & & ~~~~~~~\times (1-P(d<d_0;D_I)) \nonumber \\
& & \eeqa

Then the total inclusive rate for producing two leptons of the same
sign, summed over all three lepton flavors, is
 \beqa
\sum_{\hbox{lepton flavor}} \sigma(p+ \overline{p} \rightarrow h
\rightarrow l^{\pm} l^{\prime
 \pm}+X)_{tot} & = & 0.86 \hbox{pb}
 \sum_I Br(h \rightarrow N_I N_I)( P(d<d_0;D_I) )^2 \nonumber \\
 & & + 0.072 \hbox{pb} \sum _I Br(h \rightarrow N_I N_I)P(d<d_0;D_I)
 \nonumber \\ & &
%0.43 P(d<2 \hbox{cm};D)^2 \hbox{pb} + 0.04
%P(d<2 \hbox{cm};D) \hbox{pb}
\label{dilepton-total} \eeqa In (\ref{dilepton-total}) decays to
$\tau$ leptons were included to obtain a result independent of the
Dirac matrix elements. It therefore overestimates the number of
signal events accepted by the experimental analysis, since the
search only looked for electrons and muons (including those from
$\tau$ decays). Branching fractions for decays into charged leptons
with hadronically decaying $\tau$ not included can be computed from
Table \ref{Ffactor}, but are clearly model-dependent. The result of
such a computation is bounded from above by (\ref{dilepton-total}).

Of the right-handed neutrinos, there will be one that has the
shortest decay length and therefore largest probability factor, call
it $P_*$. The other right-handed neutrinos will have $P(d<d_0;D_I)
\leq P_*$. It is then clear that the above cross-section is
maximized when all of the right-handed neutrinos have the same
$P(d<d_0;D_I)$ or, that is, the same decay length. Then \beqa
\sum_{\hbox{lepton flavor}} \sigma( h \rightarrow l^{\pm} l^{\prime
 \pm}+X)_{tot} & \leq & \left[0.86 \hbox{pb} (P_*)^2 +
0.072 \hbox{pb} P_* \right] \sum_I Br(h \rightarrow N_I N_I )
\label{dileptonupperbound}
 \eeqa
I caution the reader that this is an upper bound that might not be
saturated in specific models.

The CDF analysis \cite{CDFsamesignlepton} requires that the
pseudorapidity of each lepton satisfy $|\eta|<1.1$ and that the
leading lepton has $E_T > 20$ GeV. These cuts will introduce an
acceptance that is crudely $1/2$ for each lepton, and $3/4$ for the
transverse energy cut, or $\approx 0.2$ in total. From
(\ref{dileptonupperbound}) with $D \approx 10$ cm and setting
$\sum_I Br(h \rightarrow N_I N_I)=1$, there are approximately 8
accepted events per 1 fb$^{-1}$. The analysis finds 44 events
consistent with 33 expected background events. Decay lengths on the
order of $10$ cm do not appear to be excluded by this search,
although a detailed studied is certainly warranted. A more careful
analysis could probably place interesting constraints for $D \leq
O($5cm$-10$cm$)$ assuming $\sum_I Br(h \rightarrow N_I N_I) \simeq
1$, or $\sum_I Br(h \rightarrow N_I N_I) \ltap O(10^{-2})$ for $D_I
\ll O($cm$)$.

In specific models the upper bound (\ref{dileptonupperbound}) may
not be saturated. For instance, if the decay lengths are
hierarchical, then only one right-handed neutrino flavor may have a
significant contribution to the sum (\ref{dilepton-total}). If so,
the overall rate is suppressed by the branching ratio for the Higgs
to decay into that state. In a minimal flavor violation scenario
where the couplings (\ref{o51}) are unsuppressed, that branching
ratio is approximately $1/3$. Then with one decay length of
approximately 10 cm there would only be roughly $3$ accepted events
per fb. If all of the decay lengths are larger than $10$ cm then the
acceptance decreases since there are fewer prompt events.

\subsection{Long-lived Neutral Particles: Displaced Muons and
Delayed Photons} \label{sec:longlivedneutrals}

The best strategy for finding the right-handed neutrinos is to look
for displaced vertices. Standard Model backgrounds are practically
nonexistent, so such events should be easy to find and interpret.
Both CDF and D0 \cite{longlivedneutralD0} have searches for a
long-lived neutral particle decaying into specific channels, such as
into a muon pair and neutrino \cite{longlivedneutralD0}, or into a
photon with jets and missing energy \cite{longliveddecaytophotons},
or to a physical $Z$ boson \cite{longliveddecaytoZ}. Although these
searches are optimized to constrain models of supersymmetry with
$R$-parity violation or with low-energy gauge mediated supersymmetry
breaking, their analysis should apply to the right-handed neutrino
decays discussed here. While it shall be seen below that no limits
yet exist, they are close. A reanalysis of their data after
optimizing the search for the specific decay modes of the
right-handed neutrinos could provide interesting limits.

Although there are several older searches, the best limits are from
the D0 search \cite{longlivedneutralD0} and the CDF search
\cite{longliveddecaytophotons}.

 Limits on the model can be improved by optimizing the search
to other decay modes having a higher branching ratio. For example,
the right-handed neutrino has a branching ratio to a muon and quark
pair that is about an order of magnitude larger than to a muon pair
and neutrino. By searching for jets and a single lepton that
reconstruct to a displaced vertex, the sensitivity to this model
should in principle be improved.

\subsubsection{Displaced Muon pairs}
\label{subsec:diplacedmuonpairs}

The DO search \cite{longlivedneutralD0} looks for a neutral particle
that travels a macroscopic distance and then decays into a neutrino
and a pair of muons having opposite signs. In particular, they
specifically look for a pair of opposite-signed muons that
reconstruct to a secondary vertex. To have a high efficiency at
reconstructing the muons, they restrict the analysis to decays that
occur within the inner region of their central detector, or
$5$cm$<d<20$ cm. Right-handed neutrinos having decay lengths outside
of this region will have a reduced acceptance and the limits will be
weaker. The following discussion implicitly assumes that all three
right-handed neutrinos have decay lengths in this region, although
the formulae allow for the more general situation.

Each event will have two right-handed neutrinos pair-produced in the
decay $h \rightarrow N_I N_I$. Moreover, at least one of these
right-handed neutrinos must decay within the signal region into a
muon pair, since both muons must come from the same vertex. The
probability for that to happen is denoted by $P^{(I)}_{displ}$ and
estimated below. The effective cross-section for producing di-muon
pairs in the signal region is then \beqa \sigma(\hbox{displaced}
~\mu^+ \mu^- +X)= \sigma(p+\overline{p} \rightarrow h +X) \sum_I
\left[Br(h \rightarrow N_I N_I) P^{(I)}_{displ} \right] ~.
\label{longlivedneutral} \eeqa Next I will estimate
$P^{(I)}_{displ}$ by splitting it into two parts.

The probability of a right-handed neutrino decaying into $\mu^+
\mu^-$ is flavor-dependent and denoted by $P^{(I)}_{\mu^+ \mu^-}$.
From Table \ref{Ffactor} and using $|D_{JI}|^2=|[m_D]_{IJ}|^2
M^{-1}_I$ this probability is given by \beqa P^{(I)}_{\mu^+ \mu^-}
&=& \frac{|[m_D]_{I2}|^2}{[m_D m^{\dagger}_D]_{II}}
\left(\frac{0.59}{N_{tot}} \right)+\left(\frac{0.13
c_Z}{N_{tot}}\right) \sum_{J \neq 2} \frac{|[m_D]_{IJ}|^2}{[m_D
m^{\dagger}_D]_{II}} \label{pIfirstline} \eeqa The first term in the
first line is due to interfering charged current and neutral
currents, and the second set of terms is due to neutral currents.
Using $N_{tot} \simeq 12$ and $c_Z \simeq 1$, an approximation to
(\ref{pIfirstline}) is \beqa
 P^{(I)}_{\mu^+ \mu^-} & \simeq&  0.05  \frac{|[m_D]_{I2}|^2}{[m_D m^{\dagger}_D]_{II}}
 + 0.01  \sum_{J \neq 2} \frac{|[m_D]_{IJ}|^2}{[m_D
m^{\dagger}_D]_{II}}
 \eeqa which is bounded from above by \beqa P^{(I)}_{\mu^+ \mu^-} &
\leq & 0.05 \frac{1}{[m_D m^{\dagger}_D]_{II}} \left[ |[m_D]_{I2}|^2
+ \sum_{J \neq 2}  |[m_D]_{IJ}|^2 \right] =0.05 ~.
 \label{pmumuupperbound} \eeqa
 A lower bound is
 \beqa
P^{(I)}_{\mu^+ \mu^-} & \geq & 0.01 \label{lowerboundPmumu} \eeqa
and is saturated only if $[m_D]_{I2}=0$. Saturation of the upper
bound occurs only if $[m_{D}]_{IJ}=0$ for $J \neq 2$. This condition
is not possible to physically realize since it requires the Dirac
mass matrix to be of rank 1 and therefore implies that two active
neutrinos are massless. Another way to say that is that this
condition requires that only $\nu_2$ has mass mixing with the
right-handed neutrinos, and therefore $\nu_1$ and $\nu_3$ are
decoupled and remain massless. Nonetheless, it is useful to have a
model-independent upper bound, albeit not possible to saturate given
the data on neutrino masses.

Next, I approximate the combined probability that at least one
right-handed neutrino in the pair decays in the signal region into a
muon pair as \beqa {\cal P}^{(I)} _{displ} &=&2 P_{I}(1-P_{I})
P^{(I)}_{\mu^+ \mu^-} + 2 P^2_I P^{(I)}_{\mu^+ \mu^-}(1-
P^{(I)}_{\mu^+ \mu^-}) + P^2_I (P^{(I)} _{\mu^+ \mu^-})^2
\nonumber \\
&=& 2 P_{I}(1-P_{I}) P^{(I)}_{\mu^+ \mu^-} + P^2_I P^{(I)}_{\mu^+
\mu^-}(2- P^{(I)}_{\mu^+ \mu^-}) \label{pdispl}~.
 \eeqa
Here $P_I$ is the probability that a right-handed neutrino $N_I$
with average decay length $D_I$ decays within the signal region. As
in previous sections, to estimate that I treat the signal region as
a sphere - rather than a
 tube - extending from $R=5$ cm out to $R=20$ cm and approximate
$P_{I} \equiv P(5 \hbox{cm},20 \hbox{ cm}; D_I)$ where \beq
P(R_1,R_2;D) \equiv e^{-R_1/D} - e^{-R_2/D}  \eeq is the probability
that a right-handed neutrino of average decay length $D$ decays a
distance between $r=R_1$ and $r=R_2$ from the primary vertex. The
first term in $P^{(I)}_{displ}$ is the probability that one
right-handed neutrino decays in the signal region into a muon pair,
the second term is the probability that both decay in the signal
region, but only one decays into a muon pair, and the third term
describes the case where both decay in the signal region into muon
pairs.

An upper bound on (\ref{longlivedneutral}) can be obtained by first
noting from (\ref{pmumuupperbound}) that $P^{(I)}_{\mu^+ \mu^-} <
0.05$. Then $P^{(I)}_{displ} \leq 0.1 P_I(1-P_I)+0.1 P^2_I<0.07$
with the maximum occurring around $D_I \simeq 10$cm. Then with
$\sigma(p+\overline{p} \rightarrow h +X) \leq 2.5$ pb, $\sum_I BR(h
\rightarrow N_I N_I) \leq 1$ and setting $D_I \simeq$10 cm for all
three right-handed neutrinos to maximize the rate, one obtains the
model-independent bound \beq
 \sigma(\hbox{displaced}~ \mu^+ \mu^- +X) < 0.175 ~\hbox{pb}~.
\label{sigmapmumuupperbound} \eeq As discussed below
(\ref{pmumuupperbound}), it is not possible to saturate this bound
with any model consistent with what is known about the active
neutrino masses. However, the lower bound \ref{lowerboundPmumu}
suggests that values might be only a factor of a few smaller in
models satisfying the assumptions listed above.

For comparison, the limits from D0 \cite{longlivedneutralD0} vary
considerably, with their best limit given by $0.14$ pb which is
slightly below the upper bound (\ref{sigmapmumuupperbound}). That
the model-independent bound is slightly larger than the best
experimental limit probably does not restrict any of the parameter
space, since as I have already stressed, the model-independent bound
cannot be saturated with any realistic model since it would require
two active neutrinos to be massless. Yet that these upper bounds are
so close to one another warrants a more careful analysis than
described here, in order to more properly determine the acceptance
of the signal in the signal region. A meaningful lower bound to the
signal rate cannot be obtained since the rate decreases rapidly once
the average decay lengths lie outside of the signal region.

\subsubsection{Displaced Photons}
\label{subsec:displacedphotons}

The CDF search \cite{longliveddecaytophotons} for displaced photons
looks at inclusive events with a photon having $E^T> 25$ GeV, at
least one jet with $E_T > 30$ GeV, and missing energy greater than
approximately 30 GeV \cite{longliveddecaytophotons}. In principle
long-lived decays of right-handed neutrinos could be constrained by
this analysis, since it has a rare decay $N_I \rightarrow \gamma
+$missing energy. The required jet can be produced from either the
decay of the other right-handed neutrino in the event, or from a
hadronically decaying electroweak gauge boson produced in
association with the Higgs boson.

Photons were required to be ``out-of-time" with the primary events
by approximately 2-10ns \cite{longliveddecaytophotons}. Whether this
occurs in a model depends on the right-handed neutrino's average
decay length and whether it is non-relativistic. Both conditions are
required \cite{longliveddecaytophotons}: a non-relativistic
right-handed neutrino that decays promptly $( \tau \ll
O(\hbox{ns}))$ will not produce a delayed photon; and a relativistic
right-handed neutrino with a long average decay length will produce
on average a photon moving in the same direction, which arrives at
the calorimeter simultaneously with other relativistic particles in
the event. Values of the underlying parameters satisfying these
conditions certainly occur. For a lifetime of
$O(2\hbox{ns}-10\hbox{ns})$ corresponds to a $c \tau$ of
approximately $O(60cm-3m)$, which occurs for a wide range of
neutrino mass parameters (see Eqn. \ref{decaylength1}).
Non-relativistic right-handed neutrinos, with say $\beta \ltap 0.6$,
are more difficult to obtain, but they do occur for masses $M_I
\gtap 0.4 m_h$. The actual region of parameters having a decent
acceptance will not be identified, but it is clear that it is not
tiny. The comments below are intended to apply to this region.

If there are other energetic, promptly decaying particles occurring
in the event, then they will arrive at the detector before the
photon. For these class of events the distribution of the time delay
for the photon is highly asymmetric, as it does not extend into
regions with negative values of the time delay $\Delta t$. This
feature was utilized in the CDF search to measure their background
to the signal. There the background to the signal, occurring from
beam-related events and Standard Model processes, was obtained by
extrapolating a measurement of the background in the region $\Delta
t <0$, where no signal is expected, into the region  $\Delta t
>0$ were a signal is expected.

This feature of the analysis raises an issue for signal events
occurring from primary Higgs production. For here all the `hard
events' are delayed. On average the decay of the two right-handed
neutrinos will be out-of-time by $O(1/\Gamma)$ with respect to each
other. But the jet used in the event selection is produced from the
decay of a right-handed neutrino, which is itself delayed. In half
of the events then, the jet will be delayed with respect to the
photon. The distribution of the photon's time delay with respect to
the arrival time of the jet will then be symmetric about zero. How
well the CDF analysis can constrain such signal events is important
to determine, since the search is beginning to cut into an
interesting region of parameter space (see below). Certainly a limit
can be set, since only a few background events were observed in the
$\Delta t<0$ region (from Fig. 1 of \cite{longliveddecaytophotons}).

This issue does not arise in associated production of the Higgs
boson, since the photon from the decay of one of the right-handed
neutrinos is delayed with respect to the particles produced from the
prompt decay of the electroweak gauge boson.

 The effective production cross-section for the signal events
is \beqa \sigma(\hbox{displaced} \gamma + \hbox{jets+missing
energy})&=& \sigma(p+ \overline{p} \rightarrow h +X)  \sum_I Br(h
\rightarrow N_I N_I) \nonumber \\
& & \times 2 Br(N_I \rightarrow \gamma +\hbox{missing energy})
\nonumber \\
& & \times Br(N_I \rightarrow \hbox{jets} +X) \eeqa As discussed in
Section \ref{subdominantdecays}, the branching fraction for the
radiative decay is approximately $7 \times 10^{-5}|c_M|^2$ and is
independent of the Dirac matrix elements if minimal flavor violation
is assumed. $c_M$ is a linear combination of a calculable one-loop
contribution and an unknown coefficient $c^{(6)}_M$ which is
determined by the coefficients of the two dimension 6 operators that
contribute to this decay. While this nominal value for the branching
fraction is too tiny to be relevant here, it increases to $0.007
|c^{(6)}_M|^2 (\hbox{TeV}/\Lambda)^4$ for $\Lambda \ltap 3 $ TeV.

Missing energy in the event occurs from the radiative decay
producing on average approximately $m_h/4$. Additional missing
energy can be obtained from the hadronic decay of the second
right-handed neutrino in the event, or from a leptonic decay of a
$W$, if present. From Table \ref{BFfordecaysintoquarks} the
branching fractions for a right-handed neutrino to decay into light
quarks or heavy quarks and missing energy totals to 0.21 and is
model-independent. The branching fraction for it to decay into jets
and any charged lepton is approximately 0.5 and also
model-independent. Although these latter decays do not produce any
missing energy, I will be conservative and include them in my
estimate below. I note that if $m_h \gtap 120$ GeV then such events
can produce on average enough missing energy from the radiative
decay alone to pass the missing energy cut. Then \beqa
\sigma(\hbox{displaced} \gamma + \hbox{jets+missing energy}) &=& 2.5
\hbox{pb} \times 2 \times 7 \times 10^{-3} |c^{(6)}_M|^2
\left(\frac{\hbox{TeV}}{\Lambda} \right)^4 \times 0.71 \nonumber
\\
& & \times \sum_I Br(h
\rightarrow N_I N_I) \nonumber \\
& = & 0.025 \hbox{pb} |c^{(6)}_M|^2 \left(\frac{\hbox{TeV}}{\Lambda}
\right)^4 \sum_I Br(h \rightarrow N_I N_I)  \eeqa where both primary
and associated Higgs production have been included. The ``2" counts
the two possibilities for a radiative decay from one of the two
right-handed neutrinos. With 100$\%$ acceptance and 570pb$^{-1}$ of
data and $\sum_I Br(h \rightarrow N_I N_I)=1$ there would have been
14 events produced for these fiducial values. The CDF analysis
imposed a number of cuts that reduces the acceptance of the signal.
For the gauge-mediated supersymmetry breaking models analyzed in
\cite{longliveddecaytophotons} the acceptance is listed in their
Table 1 as 23$\%$. Assuming a similar acceptance here implies 3
accepted events for the fiducial values above. Since only 1 event
above background was seen, a mild suppression of $|c^{(6)}_M| \simeq
0.6$ results in only 1 accepted event. With this value,
approximately 5 events would have been produced. This estimate of
the limit on $|c^{(6)}_M|$ is consistent with a conclusion of
\cite{longliveddecaytophotons} that their analysis is general enough
to exclude any model producing more than 5.5 events containing jets,
missing energy and a delayed photon with time-delay of
$O(2\hbox{ns}-10\hbox{ns})$. Increasing $\Lambda$ by a factor of 2
decreases the rate by a factor of 16, so no limit can be set for
$|c^{(6)}_M|=1$, $\Lambda \simeq 2$ TeV ($\approx 1$ event would
have been produced). In interpreting these conclusions, it is
important to recall that they only apply to those values of the mass
parameters which result in non-relativistic right-handed neutrinos.
Yet it is interesting that the search for delayed photons is
beginning to exclude $\Lambda \simeq 1 ~\hbox{TeV}$ if the dominant
decay of the Higgs boson is to non-relativistic right-handed
neutrinos.

Further analysis of these decays to optimize the experimental search
is certainly warranted. Moreover, it is important to determine the
sensitivity of the search strategy to delayed photons occurring from
primary Higgs production, for the reasons already discussed above.
If this is indeed an issue, then the constraint obtained above may
have been too cautious.

\subsection{Inclusive Searches for Anomalous Photons}

Signature-based searches
\cite{CDFpublicnotephotonphoton,CDFpublicnotephotonlepton}involving
photons are of several types. There are inclusive searches for $
\gamma \gamma +X$ where $X=\gamma$, missing energy, or a prompt
lepton. There are also inclusive searches for $\gamma l $ and
$\gamma l l $. These searches may potentially constrain the model,
since photons may be produced from both real and fake physics
sources.

As discussed in Section \ref{subdominantdecays}, additional
operators at dimension 6 introduce a small branching fraction for
right-handed neutrinos to decay \beq N_I \rightarrow \gamma \nu_L,
\gamma \overline{\nu}_L ~. \eeq Since the right-handed neutrinos are
displaced a macroscopic distance on average, these photons will not
point back to the primary vertex. From (\ref{Nnugammadecay}), the
inclusive branching fraction for this decay is approximately \beqa
Br(N_I \rightarrow \gamma +\hbox{missing energy}) &=& 7 \times
10^{-5}
|c_M|^2 \nonumber \\
& \simeq &  |c^{(6)}_M|^2 \left(\frac{\hbox{TeV}}{\Lambda}\right)^4
~. \eeqa The last relation occurs for $\Lambda \ltap 4 \pi v \simeq
3$ TeV, which will be assumed for the remainder of this subsection.

 To estimate the event rate, note that summing over
the three dominant production mechanisms for the Higgs boson gives
$\sigma_h \simeq 2.5$ pb for $m_h \simeq 115$ GeV . Tnen the
inclusive rate to produce a photon from one of the two right-handed
neutrinos is  \beq \sigma(h \rightarrow \gamma +\hbox{missing
energy} +X) \simeq 35 \sum _I Br(h \rightarrow N_I N_I)
|c^{(6)}_M|^2 (\hbox{TeV}/\Lambda)^4 ~ \hbox{fb} \eeq where $X$ is
not yet specified. The energy of the photon is $\approx m_h /4
\approx 25$ GeV and the missing energy is $\approx m_h/4 \approx 25$
GeV plus whatever is lost in $X$.

The largest rate occurs from primary Higgs production. From Table
\ref{BFfordecaysintoquarks} the second right-handed neutrino will
decay into charged leptons plus jets 50$\%$ of the time, and into
two charged leptons and missing energy about $24 \%$ of the time. As
these leptons must be prompt in order to be included in the
analysis, such events are suppressed by a factor of $P(d<d_0=\hbox{2
cm};D_I)$. Leptons from decays of a $W$ produced in association with
the Higgs are prompt, but the overall rate is lower. Adding both
production mechanisms gives \beqa \sigma(h \rightarrow \gamma
+l+\hbox{missing energy} +X) & \simeq & \sum_I Br(h \rightarrow N_I
N_I)\left[ 35 P(d<d_0;D_I)(0.75) \right. \nonumber
 \\ & & \left. +(0.007)(300)(0.2) \right] \nonumber
\\
& & \times |c^{(6)}_M|^2 (\hbox{TeV}/\Lambda)^4 ~ \hbox{fb}
\nonumber \\ & =& \sum_I Br(h \rightarrow N_I N_I) \left[ 26
P(d<d_0;D_I) + 0.4 \right]  \nonumber \\
& & \times |c^{(6)}_M|^2 (\hbox{TeV}/\Lambda)^4 ~ \hbox{fb}
\label{glevents} \eeqa With $D_I > 10$ cm,  $P(d< \hbox{2 cm};D_I)
\ltap 0.2$ and setting $\sum_I Br(h \rightarrow N_I N_I)=1$ gives
approximately $6 |c^{(6)}_M|^2 (\hbox{TeV}/\Lambda)^4$ events
produced per fb$^{-1}$. The number of events detected is reduced by
the acceptance which I have not included here. From inspecting
Figures 2,3 and 4 from the CDF Public Note
\cite{CDFpublicnotephotonlepton} giving distributions of lepton plus
photon plus missing energy events, a handful ($\approx 13$ when
combined with (\ref{glfakeevents}) below) of such events do not
appear to be excluded, although a more careful analysis is
warranted. Limits can probably be set for smaller values of the
average decay length. In particular, for $D_I \ll O(5 \hbox{cm})$
limits $\sum_I Br(h \rightarrow N_I N_I) \ltap 0.1$ could probably
be set assuming $c^{(6)}_M=1$ and $\Lambda =1 $TeV.

In principle one might consider signal events containing two photons
and missing energy $\simeq m_h/2 \simeq 60$ GeV. Then \beq \sigma(h
\rightarrow \gamma \gamma +\hbox{missing energy} +X) \simeq 0.2 \sum
_I Br(h \rightarrow N_I N_I) |c^{(6)}_M|^2 (\hbox{TeV}/\Lambda)^4 ~
\hbox{fb} \eeq which is unfortunately too small.

 A potential source for fake photons
occurs if a right-handed neutrino decays inside the electromagnetic
calorimeter into a final state containing an electron. These
electrons will be contained in the calorimeters and so will not
produce a track. How this electron is identified depends on the
analysis. Several Tevatron analyses explicitly state that a fully
contained energy deposition in the electromagnetic calorimeter
without an associated track will be identified as a photon. While
such an electron is in principle distinguishable from a photon and
might be misidentified as a neutral hadron, to be conservative in
ascertaining whether these searches may have ``caught" some of the
signal events, I will assume that some of these electrons are
misidentified as photons. This fraction will be denoted by
$f_{\gamma |e}$ and I will not try to estimate it.

The electromagnetic calorimeter is shaped like a cylinder and not at
a fixed distance from the primary vertex. It is therefore sensitive
to a range of average decay lengths. Since the calorimeter is not
big, the fraction of decays occurring in that region is suppressed.
The CDF calorimeter has a width of about 30cm and extends from
approximately $R_1=1.7$m out to approximately $R_2=2$m. The fraction
of events decaying in this volume depends sensitively on the
geometry of the calorimeter and requires a detector simulation. To
obtain a rough estimate, I will crudely approximate the
electromagnetic calorimeter as a hollow sphere with radius extending
between 1.7m and 2m. Then the fraction of right-handed neutrinos of
average decay length $D_I$ that decay in this region is \beq P
_{\gamma}(D_I) \equiv P(R_1<d<R_2;D_I) ~. \eeq

Let me begin with the $``\gamma \gamma" +$missing energy events from
primary Higgs production. Events with missing energy and two fake
photons occur from $h \rightarrow N_I N_I $ where either one or both
right-handed neutrinos decay inside the electromagnetic calorimeter.
If only one decays inside, then $N_I \rightarrow e^+ e^-+$missing
energy with branching fraction $P^{(I)}_{e^+ e^-}$. One can show
$P^{(I)}_{e^+ e^-} \leq 0.05$ using reasoning similar to that found
in Section \ref{subsec:diplacedmuonpairs} to derive an identical
upper bound on decays to muon pairs. For decays to electrons,
saturation of the bound requires $[m_D]_{IJ}=0$ for $J \neq 1$ which
as discussed in Section \ref{subsec:diplacedmuonpairs} is
unrealistic. If both right-handed neutrinos decay in the
electromagnetic calorimeter then the majority of the fake events
occur when each decays to an electron. Although the dominant decay
of $N_I$ is into charged leptons and light quarks and occurs with
branching fraction of 1/2, both right-handed neutrinos cannot decay
into this channel since there is no missing energy.  One $N_I$ must
decay to $e+l+$missing energy, whose branching ratio $P^{(I)}_{e l}$
is less than 0.24. Then
\begin{eqnarray}
\sigma(h \rightarrow ``\gamma \gamma"+X) & =& 2.5 \hbox{pb} \sum_I
Br(h \rightarrow N_I N_I) \nonumber \\
& & \times f^2_{\gamma |e} \left[ 2P
_{\gamma}(D_I)(1-P_{\gamma}(D_I))P^{(I)}_{e^+ e^-} +  2
P^{(I)}_{eqq} P^{(I)}_{e l} (P_{\gamma} (D_I))^2 \right] \eeqa
Varying $D_I$ to maximize the signal gives
$P_{\gamma}(D_I)=P(1.7<d<2;D_I) \leq 0.06$ and $2P
_{\gamma}(D_I)(1-P_{\gamma}(D_I)) \leq 0.12$. Using $P^{(I)}_{eqq}
\leq 0.5$ and $P_{el} \leq 0.24$ one obtains
 finds \beqa \sigma(h
\rightarrow ``\gamma \gamma"+X) &<& 2.5 \hbox{pb} f^2_{\gamma |e}
\sum_I Br(h \rightarrow N_I N_I) \left[ (0.12)(0.05) + 0.25
(0.06)^2 \right] \nonumber \\
&=& 0.017 \hbox{pb} f^2_{\gamma |e} \sum_I Br(h \rightarrow N_I N_I)
\eeqa
%The upper bound is saturated only if all three right-handed
%neutrinos have average decay lengths $D_I \simeq O(2m)$ and
%$[m_D]_{IJ}=0$ for $J \neq 1$. As discussed in Section
%\ref{subsec:diplacedmuonpairs}, this latter assumption is not
%realistic since it implies two massless active neutrinos.
A more accurate estimate of the rate can be determined in specific
models, but it will depend on the Dirac mass matrix elements. It
will be bounded by the result above, but will probably only be a
factor of a few smaller.

The number of events detected is reduced from the above result by
several factors. First, there is the unknown probability $f_{\gamma
| e}$ of misidentifying a trackless electron produced in the
electromagnetic calorimeter as a photon. Next, the CDF  analysis
\cite{CDFpublicnotephotonphoton} requires that both photons have
pseduo-rapidity $|\eta|<1$, which lowers the acceptance. Finally,
the acceptance is further reduced because the events described above
have little missing energy, since most only have one neutrino that
on averages has $E \simeq m_h/6$. Assuming $100\%$ acceptance for
the misidentification of electrons $(f_{\gamma |e}=1)$ and $\sum_I
Br(h \rightarrow N_I N_I)=1$, there are less than $\approx 17
$fb$^{-1}$ produced events. Inspecting their Figures 1 and 2, in
their control region they expected 115 events with missing $E_T> 20$
GeV, and observed 126. In their signal region they expected 174 and
saw 196 events. So no constraint currently exists from this channel.

Events with a larger amount of missing energy require associated
production of $hW$ or $hZ$, where the gauge bosons decay to
neutrinos. Then the analysis above mostly carries over, except that
now if both right-handed neutrinos decay inside the electromagnetic
calorimeter both can decay to a charged lepton and quarks, since no
missing energy is needed from the right-handed neutrinos. Then with
$Br(N_I \rightarrow e+ X) \equiv P^{(I)}_{e} \leq 0.7$,
\begin{eqnarray}
\sigma(h \rightarrow ``\gamma \gamma"+X)_{assc prod} &=& (0.3)(0.3)
f^2_{\gamma |e} \hbox{pb} \sum_I Br (h \rightarrow N_I N_I) \nonumber \\
& & \times \left[ 2P
_{\gamma}(D_I)(1-P_{\gamma}(D_I))P^{(I)}_{e^+e^-} + P^{(I)}_{e}
P^{(I)}_{e} (P_{\gamma} (D_I))^2 \right]  \nonumber \\ & \leq & 0.7
\hbox{fb} f^2_{\gamma |e} \sum_I Br (h \rightarrow N_I N_I)
\end{eqnarray}

Next consider $ \gamma l +X$ events where an electron produced in
the electromagnetic calorimeter is faking a photon. The lepton must
be prompt and can be produced from the other right-handed neutrino
decaying leptonically in the acceptance region $d < d_0 \simeq 2$
cm, or from the leptonic decay of a $W$. The former events are
suppressed by the additional factor of $P(d<\hbox{2 cm};D_I)$, and
the latter events by the lower production cross-section. In total
\begin{eqnarray} \sigma(p+ \overline{p} \rightarrow h +Y \rightarrow l ``\gamma" +X) &=&
\sum_I  Br(h \rightarrow N_I N_I) \left[(2) 2.5 \hbox{pb}
P(d<d_0,D_I) P_{\gamma}(D_I) P^{(I)}_{e} f_{\gamma} \right.
\nonumber \\ &  & \left.
 +
0.3 \hbox{pb} (0.2) P_{\gamma}(D_I) P^{(I)}_e
f_{\gamma}\right] \nonumber \\
   &<& 7 f_{\gamma} ~\hbox{fb}  \sum_I  Br(h \rightarrow N_I N_I)
   \nonumber \label{glfakeevents} \eeqa where $P(d<2
\hbox{cm},D_I)\times P_{\gamma}(D_I) <1.2\times 10^{-3}$,
$P_{\gamma}(D_I) < 0.06$ and $P^{(I)}_e <0.75$ have been used. With
1 fb$^{-1}$ this leads to a handful of events produced. Inspecting
Figures 2 and 3 in \cite{CDFpublicnotephotonlepton} giving the
distribution for events in the $e \gamma$ and $\mu \gamma$ missing
energy samples, it appears that the addition of a handful of events
produced (when combined with (\ref{glevents})) in each channel is
not excluded.

\section{Beyond Minimal Flavor Violation}
\label{sec:beyondmfv}

 The minimal flavor violation hypothesis
\cite{mfv} has been used throughout since dimension 6 operators
involving {\em quarks} or {\em leptons} must be suppressed in order
to be consistent with flavor changing neutral current processes.
This hypothesis seems too restrictive however, for operators that
involve only right-handed neutrinos, since they are more poorly
constrained. In particular, there are no experimental constraints on
the operator \beq \frac{c_{IJ}}{\Lambda} N_I N_J H^{\dagger} H \eeq
(assuming $\Lambda
> O(m_R)$ for consistency of the effective theory).

With minimal flavor violation, the couplings $c_{IJ}$ are aligned
with the right-handed neutrino masses $m_R$, so that they are
simultaneously diagonalizable. In that case the couplings to the
Higgs boson are diagonal in the right-handed neutrino mass basis.

In a more general context though with arbitrary $c_{IJ}$ not aligned
with the right-handed neutrino masses, non-universal couplings can
occur and they have interesting consequences. For now the couplings
to the Higgs boson are \beq \frac{v}{\Lambda} c^{\prime}_{IJ} N_I
N_J h \eeq with $c^{\prime} _{IJ} \neq \delta _{IJ}$. Now heavier
right-handed neutrinos can decay into lighter right-handed neutrinos
through an off-shell $h$. This leads to the following new decays
\footnote{The author thanks Scott Thomas for this observation.} \beq
N_I \rightarrow N_J h^* \rightarrow N_J b \overline{b} \eeq where
$M_I > M_J$. These decays are to a two-body final state and are only
suppressed by the Higgs boson mass and the bottom Yukawa coupling.
They will dominate over the charged current and neutral current
decay processes for $ c^{\prime} \lambda _b \gtap O(\lambda_{\nu})$
which occurs for a wide range of $c^{\prime}$'s.

One then has the following scenario. The heavier two right-handed
neutrinos can decay to a bottom quark pair and a lighter
right-handed neutrino. Depending on the relative sizes of the
couplings $c^{\prime}$, the heaviest right-handed neutrino $(N_H)$
may decay preferentially to the second heaviest right-handed
neutrino $(N_M)$, which then decays to the lightest right-handed
neutrino $(N_L)$ and another bottom quark pair. The lightest
right-handed neutrino is kinematic forbidden to have these decays,
so it will decay through the charged and neutral current processes.
One could then have \beqa N_H
& \rightarrow & 2 N_L + b \overline{b} + b \overline{b} \nonumber \\
& \rightarrow & l + qq^{\prime} + b \overline{b} + b \overline{b}
\eeqa which has 7 particles in the final state.  The other decay
process $N_H \rightarrow N_L + b \overline{b}$ could occur as well.

The couplings of the Higgs boson are now non-universal, so decays to
all 6 possible pairing $N_I N_J$ occur. Decays of the Higgs boson
into right-handed neutrino can produce anywhere from 4 and 6
particles in the case of Higgs boson decays to $N_L N_L$,  to up to
14 particles in the case of decays to $N_H N_H$.

The phenomenology of this scenario is quite rich and the analysis of
the experimental constraints somewhat more involved from what I
considered in previous Sections. Because of the large number of $b$
quarks in these events they probably would have been seen at LEP, so
a limit of $m_h \gtap 114$ GeV can probably be set. The analyzes of
the searches considered in previous Sections can be carried over to
Higgs boson decays to $N_L N_L$, since the decays of the lightest
right-handed neutrinos remains unchanged. However, the rates will be
reduced by the branching fraction for the Higgs boson to decay
exclusively to the two lightest right-handed neutrinos.  The large
number particles and of bottom quarks in the other decays will make
it hard to find at hadron colliders.

\section{Conclusion}

I have discussed experimental limits on Higgs boson decays into
right-handed neutrinos within the context of minimal flavor
violation. These decays are interesting, since there is a wide range
of scales $\Lambda$ and Higgs mass in which they can be the dominant
decay mode of the Higgs boson. To set precise limits however
requires determining the acceptances for these decays, and exploring
the parameter space of the model, and I have done neither. The
parameter space for these models is large, having in addition 18
neutrino parameters. Setting limits is therefore model-dependent,
since branching fractions for right-handed neutrinos decaying into
specific final states are highly model-dependent. Model-independent
branching fractions for inclusive decays can be derived (Table
\ref{BFfordecaysintoquarks}) and are quite useful in obtaining
model-independent upper bounds to the number of events produced. I
used these upper bounds to infer limits that might occur in an
actual model, although that overestimates the rate.

Quite remarkably, much of that parameter spaces does not appear to
be excluded. Dominant decays of the Higgs boson into right-handed
neutrinos having average decay lengths larger than $O(10cm)$ appear
to be allowed. Constraints from a number of searches do exist for
average right-handed neutrino decays lengths of $O(2-5cm)$ and
smaller, and probably require that the branching fraction for the
Higgs boson to decay into such states is $O(10^{-2}-10^{-1})$. That
the branching ratios are that small may naturally occur in specific
models. For a shorter average decay length requires that the
right-handed neutrino is heavier, with average decay lengths less
than $O(cm)$ requiring that it be heavier than the $W$. Decays of
the Higgs boson into these states may naturally be small because: i)
the Higgs boson is light and this channel is not kinematically
accessible, although decays to other right-handed neutrino pairs may
be; or ii) the Higgs boson can decay into it, but then it may be
heavy enough to decay into electroweak gauge bosons, in which case
the branching fraction of the Higgs boson to decay into right-handed
neutrinos is naturally small enough for $\Lambda \simeq
O(5\hbox{TeV}-7 \hbox{TeV})$ (compare Figure \ref{decaylength1} with
Figure \ref{fig:hdecayWW}a.). Current searches are beginning to
probe regions of decay lengths larger than $O(10cm)$. These findings
justify further studies to better determine the acceptances for
these decays, and to optimize search strategies.

\section{Acknowledgements}
The author thanks Todd Adams, Michael Barnett, Jim Branson, John
Conway, Michael Dine, Paddy Fox, Ian Hinchliffe, David Kirkby, Amit
Lath, Sunil Somalwar, Matt Strassler, Scott Thomas, Chris Tully, and
Lian-tao Wang for comments, discussions and suggestions. This work
is supported by the U.S. Department of Energy under contract No.
DE-FG03-92ER40689.

\appendix{\section{Minimal Flavor Violation and Dimension 6 Operators}
\label{sec:appA}

A predictive framework for the flavor structure of the higher
dimension operators is provided by the minimal flavor violation
hypothesis \cite{mfv,mfvquarks,mfvleptons}. This hypothesis
postulates a flavor symmetry assumed to be broken by a minimal set
of non-dynamical fields or spurions, whose vevs determine the
renormalizable Yukawa couplings and masses that violate the flavor
symmetry. With the assumption of a single spurion breaking each
flavor group, the flavor structure of the higher dimension operators
is fixed in terms of appropriate powers of the spurions, or in other
words, in terms of the low-energy Yukawa couplings. Limits on
operators in the quark sector are $5-10$ TeV \cite{mfvquarks}, but
weak in the lepton sector unless the neutrinos couplings are not
much less than order unity \cite{mfvleptons}\cite{mfvcg}.

If the assumption of minimality is relaxed then limits on the scale
of these operators would be much higher. For if there are multiple
spurions breaking the same flavor group, the linear combination of
spurions appearing in a higher dimension operator does not have to
be the same combination determining the Yukawa couplings. If so, the
coefficients of the higher dimension operator cannot be expressed in
terms of the Yukawa couplings. This is dangerous, since barring
accidental cancelations, the misalignment between the higher
dimension operators and the Yukawa couplings leads to large flavor
violation. To avoid this potential problem, a minimal field content
will be assumed \footnote{Multiple fields generating right-handed
neutrinos masses are probably allowed, but not explored here.}.

The flavor symmetry in the lepton sector is taken to be
\begin{eqnarray} && G_N \times SU(3)_{L} \times
SU(3)_{e^c}\times U(1)
\end{eqnarray}
where $U(1)$ is the usual overall lepton number acting on the
Standard Model leptons. With right-handed neutrinos present there is
an ambiguity over what flavor group to choose for the right-handed
neutrinos, and what charge to assign them under the $U(1)$. In fact,
since there is always an overall lepton number symmetry unless both
the Majorana masses and the neutrino coupling are non-vanishing,
there is a maximum of two such $U(1)$ symmetries.

Two possibilities are considered for the flavor group of the
right-handed neutrinos: \beq G_N = SU(3) \times U(1)^{\prime}
~\hbox{or}~ SO(3) ~. \eeq The former choice corresponds to the
maximal flavor group, whereas the latter is chosen to allow for a
large coupling for the operator (\ref{o51}), shown below. The fields
transform under the flavor group $SU(3) \times SU(3)_L \times
SU(3)_{e^c} \times U(1)^{\prime} \times U(1)$ as
\begin{eqnarray}
N & \rightarrow & ({\bf 3},{\bf 1}, {\bf 1})_{({\bf 1},{\bf 0})} \\
L & \rightarrow &  ({\bf 1},{\bf 3}, {\bf 1})_{({\bf -1},{\bf 1})}\\
e^c & \rightarrow & ({\bf 1},{\bf 1}, {\bf 3})_{({\bf 1},{\bf -1})}
~.
\end{eqnarray}
Thus $U(1)^{\prime}$ is a lepton number acting on the right-handed
neutrinos and Standard Model leptons and is broken only by the
Majorana masses. $U(1)$ is a lepton number acting only on the
Standard Model leptons and is only broken by the neutrino couplings.
Promoting the masses and Yukawa couplings of the theory to spurions
gives for $G_N=SU(3)\times U(1)^{\prime}$,
\begin{eqnarray}
\lambda_{\nu} & \rightarrow & ({\bf \overline{3}},
\bf{\overline{3}},\bf{1})_{({\bf 0},{\bf -1})}   \\
\lambda_l & \rightarrow & (\bf{1},\bf{\overline{3}},
\bf{\overline{3}})_{({\bf 0},{\bf 0})} \\
m_R & \rightarrow & ({\bf \overline{6}}, \bf{1},\bf{1})_{({\bf
-2},{\bf 0})}~.
\end{eqnarray}

For $G_N=SO(3)$ there are several differences. First, the ${\bf
\overline{3}}$'s of $SU(3)$ simply become ${\bf 3}$'s of $SO(3)$.
Next, the $U(1)$ charge assignments remain but there is no
$U(1)^{\prime}$ symmetry. Finally, a minimal field content is
assumed throughout, implying that for $G_N=SO(3)$ $m_R \sim {\bf 6}$
is real.

As I discuss in \cite{mg}, the sizes of the coefficents of the
dimension 5 operators can now be estimated. There are several
operators at dimension 5, but only (\ref{o51}) is potentially
relevant to decays of the Higgs boson or right-handed neutrinos. All
other operators are suppressed by powers of the neutrino couplings
relative to the leading order decays. For (\ref{o51}) one finds that
its coefficients depend on the choice of flavor group. This isn't
surprising, since the operator $N_I N_I$ violates $SU(3)$, but
preserves an $SO(3)$ symmetry. One finds
\begin{eqnarray}
G_N = SU(3)\times U(1)^{\prime} &:& c  \sim a_1
\frac{{m_R}}{\Lambda} + a_2 \frac{{m_R  \hbox{Tr}[m^{\dagger}_R
m_R]}}{\Lambda^2} +\cdots
 \nonumber \\
G_N= SO(3) &:& c  \sim {\bf 1} + b_1 \frac{m_R}{\Lambda} + b_2
\frac{{m_R \cdot m_R}}{\Lambda^2} + \cdots + d_1 \lambda_{\nu}
\lambda^{\dagger}_{\nu} + \cdots \label{gnexpression}
\end{eqnarray} where $\cdots$ denotes higher powers in $m_R$ and
$\lambda_{\nu} \lambda^{\dagger}_{\nu}$. Comparing the expressions
in (\ref{gnexpression}), the only important difference between the
two is that ${\bf 1}$ is invariant under $SO(3)$, but not under
$SU(3)$ or $U(1)^{\prime}$. This is a key difference that has
important consequences for the decay rate of the Higgs boson into
right-handed neutrinos, as can be seen from Figures
\ref{fig:hdecaybb} and \ref{fig:hdecayWW}, and discussed further in
\cite{mg}.

At dimension 6 there are a number of operators. Here I focus on
those involving at least one right-handed neutrino.  One finds that
to leading order in the Yukawa couplings
\begin{eqnarray}
{\cal O}^{(6)}_1 &=& g_1 c^{(6)}_1 (N
\lambda_{\nu} \sigma^{\mu \nu} L) \widetilde{H} B_{\mu \nu} \\
{\cal O}^{(6)}_2 &=& g_2 c^{(6)}_2 (N \lambda_{\nu}
\sigma^{\mu \nu} \widetilde{H}^{\dagger} \tau ^a L)  W^a_{\mu \nu} \\
{\cal O}^{(6)} _3 &=& c^{(6)}_3 (N \lambda_{\nu}
L)( L  \lambda_l e^c)  \\
{\cal O}^{(6)} _4 &=& c^{(6)}_5 (\overline{N}
\lambda^*_{\nu} \lambda_l \gamma^{\mu} e^c )(H D_{\mu} H) \\
{\cal O}^{(6)} _5 &=& c^{(6)}_6   \partial^{\mu}
N \lambda_{\nu} L D_{\mu} \tilde{H} \\
{\cal O}^{(6)} _6 &=& c^{(6)}_7  (N \lambda_{\nu} L \widetilde{H})(
H^{\dagger} H ) \\
{\cal O}^{(6)}_7 &=& c^{(6)}_8 \overline{N} \lambda^*_{\nu}
\gamma^{\mu} \lambda_{\nu} N H^{\dagger} D_{\mu} H
\end{eqnarray}
where all of the $c^{(6)}_i$'s are numbers and not matrix-valued.
Also, $\widetilde{H} \equiv i \tau_2 H^*$. Other operators are
equivalent to those listed above either by an integration by parts
in the action, or by using the free equations of motion.

Most of these operators introduce new decay modes for the
right-handed neutrinos. But with the exception of a couple of
operators discussed below, their amplitudes are parametrically
suppressed by additional powers of Yukawa couplings compared to the
previously discussed amplitudes created by mass mixing of the
right-handed neutrinos with the left-handed neutrinos. For recall
that the latter amplitudes are $O(\lambda_{\nu})$. By inspection,
all of the dimension 6 operators listed above contain one factor of
the neutrino Yukawa coupling, a consequence of the fact that the
$U(1)$ lepton number is only violated by the neutrino couplings.
Some occur with an additional power of a Yukawa coupling.

As a result, most of these operators are irrelevant to the decay of
the right-handed neutrinos. For example, ${\cal O}^{(6)}_3$, ${\cal
O}^{(6)}_4$ and ${\cal O}^{(6)}_7$ can all be ignored since they are
suppressed relative to the other operators by at least an additional
factor of the lepton or neutrino Yukawa couplings. After electroweak
symmetry breaking, the only effect of ${\cal O}^{(6)}_6$ is to
correct the overall scale of the neutrino coupling by a small
amount. It does not contribute to the right-handed neutrino decay
since $m_h > m_R$ is assumed throughout. Operator ${\cal O}^{(6)}_7$
contributes to the decay of a Higgs boson, but not to the decay of a
right-handed neutrino. However, the contribution of this operator to
$h$ decay is suppressed by two powers of the neutrino Yukawa
coupling in the amplitude and therefore irrelevant.

Of the remaining three operators relevant to the decay of a
right-handed neutrino, ${\cal O}^{(6)}_1$, ${\cal O}^{(6)}_2$ and
${\cal O}^{(6)}_5$ contribute to $N \rightarrow l W^*$ and $N
\rightarrow \nu_L Z^*$ decays at $O(\lambda_{\nu})$ in the
amplitude.  This is parametrically the same order as decays caused
by neutral and charged current interactions. But since the dimension
6 operators are suppressed by $\Lambda^2$, contributions of ${\cal
O}^{(6)}_1$, ${\cal O}^{(6)}_2$ ${\cal O}^{(6)}_6$ to these decay
processes can be neglected.

However, both ${\cal O}^{(6)}_1$ and ${\cal O}^{(6)}_2$ contribute
to a new decay process. After electroweak symmetry breaking these
two operators generate the magnetic moment operator \beq
 e c_M \frac{1}{\Lambda^2}(N
m_D \sigma^{\mu \nu} \nu_L) F_{\mu \nu} \eeq where $F_{\mu \nu}$ is
the electromagnetic field strength and $c_M$ is a linear combination
of $c_1$ and $c_2$. This operator introduces a new decay mode \beq N
\rightarrow \gamma \nu_L~,~\gamma \overline{\nu}_L ~.\eeq This decay
is further discussed in Sections \ref{subdominantdecays} and
\ref{sec:tevatron}.

\section{Amplitudes and Decay Rates}
\label{sec:appB}

Through mass-mixing with the left-handed neutrinos $\nu_I$, the
right-handed neutrinos $N_I$ acquires the following couplings to
massive gauge bosons and leptons, \beqa {\cal L}_{N} &=&
i\frac{g}{\sqrt{2}} D_{JI} W_{\mu}^- \overline{l}_J
\overline{\sigma}^{\mu} N_I + i \frac{g g^{(\nu)} _L}{\cos \theta_W}
D_{JI} g^{(i)} Z_{\mu} \overline{\nu}_J \overline{\sigma}^{\mu} N_I
+h.c. \eeqa where the ``left-handed" and ``right-handed" fermion
couplings to the $Z$ are
\beqa g^{(i)}_L &=& T^{(i)}_3 -Q^{(i)} \sin ^2 \theta_W \nonumber \\
g^{(i)}_R & =& - Q^{(i)} \sin^2 \theta _W \eeqa and where $\sin^2
\theta_W \simeq 0.2312$ is the Weinberg angle, $T^{(i)}_3= \pm 1/2$
is the weak isospin , $Q^{(l)}=-1$, $D_{JI}= [m^T_D]_{JI} M^{-1}_I$
is the left-right mixing angle, and two-component notation is used.

For the charged-current mediated decay $N_I \rightarrow l_J q
\overline{q}^{\prime}$ one obtains the amplitude \beq A_{CC} = -i
\left(\frac{g}{\sqrt{2}}\right)^2 \left[\overline{u}(p_l)
\overline{\sigma}^{\mu} \mu(p_N) \right] \left[ \overline{u}(p_q)
\overline{\sigma}_{\mu} \nu(p_{q^{\prime}}) \right]
\frac{D_{JI}}{(p_N-p_l)^2 - m_W^2 + i \Gamma_W m_W} \eeq The
spin-averaged matrix element is \beq \frac{1}{2} \sum_{\hbox{spin}}
|{\cal A}_{CC}|^2 = |D_{JI}|^2 \frac{32 N_c}{v^4} \frac{1}{(1-
\frac{M_I^2}{m_W^2} + 2 \frac{E_l M_I}{m^2_W})^2+
\frac{\Gamma^2_W}{m^2_W}} (p_N \cdot p_q) (p_l \cdot p_{q^{\prime}})
\eeq The integral over phase space is the same as in muon decay and
be done following \cite{dynamicsSM}. This leads to the exclusive
decay rate \beq \Gamma[N_I \rightarrow l_J q \overline{q}^{\prime}]=
\frac{G^2_F M^5_I}{192 \pi^3} |D_{JI}|^2  N_c c_W  \eeq where the
factor $c_W$ is due to the final integration over the lepton energy
and is given by \beq c_G(x_G,y_G) = 2 \int^1 _0 dz z^2(3-2z)
\left((1-(1-z)x_G)^2+ y_G \right)^{-1} \label{cG} \eeq where $x_G =
M^2_I/m^2_G$, $y_G = \Gamma^2_G/m^2_G$, $c_G(0,0)=1/(1+y_G) \simeq
1$. The non-vanishing momentum transfer enhances the decay rate by
approximately $10\%$ for $m_R$ masses around  $30 \hbox{GeV}$, by
approximately $50\%$ for masses around $50$ GeV. By including the
effect of the finite width of the gauge boson this formula can also
be used in the region $M_I
> m_W$ where the $W$ boson is on-shell.

From the result above one readily obtains \beq \Gamma[N_I
\rightarrow l_J \overline{l}_{K \neq J} \nu_K ] =\frac{G^2_F
M^5_I}{192 \pi^3} |D_{JI}|^2  c_W  \eeq the only difference being
the absence of a color factor. Decays into same flavor leptons,
$K=J$, are dealt with below since there is quantum interference
between the charged and neutral currents which must be computed more
carefully.

Decays of $N_I$ to $\nu_J q_K \overline{q}_K$, $\nu_J l_{K \neq J}
\overline{l}_{K \neq J}$ or to $\nu_J \nu_{K \neq J}
\overline{\nu}_{K \neq J}$ are similarly obtained. The only
difference is to replace $N_c c_W \rightarrow \left[(g^{(K)}_L)^2 +
(g^{(K)}_R)^2 \right] N^{(K)}_c c_Z$.  Then one finds \beq
\Gamma[N_I \rightarrow \nu_J f_{K \neq J} \overline{f}_{K \neq J} ]=
\frac{G^2_F M^5_I}{192 \pi^3} |D_{JI}|^2 \left[(g^{(K)}_L)^2 +
(g^{(K)}_R)^2 \right] N^{(K)}_c c_Z  \eeq

 The amplitude for $N_I
\rightarrow l^-_J l^+_J \nu_J$ has interfering contributions from
charged and neutral current interactions. The computation of the
amplitude and tricky factors of $(-1)$ etc., are similar to the
amplitudes for the process $\overline{\nu} e \rightarrow
\overline{\nu} e$ which is described in, for example
\cite{ecommins}. In what follows I parallel their discussion. The
amplitude for the charged current contribution is
 \beqa
{\cal A}_{CC}&=& -i D_{JI}\left(\frac{g}{\sqrt{2}}\right)^2
\left[\overline{u}(p_{l^-}) \overline{\sigma}^{\mu} u(p_N)\right]
\left[\overline{u}(p_{\nu}) \overline{\sigma}_{\mu}
\nu(p_{l^+})\right]
\frac{1}{(p_N-p_{l^-})^2-m^2_W} \nonumber \\
&=& + i D_{JI}\left(\frac{g}{\sqrt{2}}\right)^2
\left[\overline{u}(p_{l^-}) \overline{\sigma}^{\mu} \nu(p_{l^+})
\right] \left[\overline{u}(p_{\nu}) \overline{\sigma}_{\mu}
u(p_N)\right] \frac{1}{(p_N-p_{l^-})^2-m^2_W} \eeqa where a Fierz
transformation was performed in the last step. Two-component
notation is used here. In ignoring the lepton mass, the coupling of
the $Z$ to the left-handed and right-handed lepton currents do not
interfere in the overall rate, so I can consider them separately.
Only the contribution of the left-handed lepton neutral current to
the amplitude interferes with the charged current amplitude. One
finds \beqa {\cal A}^{(L)}_{NC} &=& i D_{JI} \frac{g^2}{2}
\frac{g^{(l)}_L}{\cos^2 \theta_W} \left[\overline{u}(p_{l^-})
\overline{\sigma}^{\mu} \nu(p_{l^+})\right]
\left[\overline{u}(p_{\nu}) \overline{\sigma}_{\mu} u(p_N)\right]
\frac{1}{(p_N-p_{\nu})^2-m^2_Z} \eeqa In the first line of the
charged current amplitude there is an overall $(-)$ sign relative to
the neutral current amplitude ${\cal A}^{(L)}_{NC}$ because of a
relative ordering of the anti-commuting creation and annihilation
operators that occurs when computing the amplitude. The total is
\beqa {\cal A}_{CC} + {\cal A}^{(L)}_{NC} &=& -i \frac{g^2}{2}
D_{JI} \left[\overline{u}(p_{l^-}) \overline{\sigma}^{\mu}
\nu(p_{l^+})\right] \left[\overline{u}(p_{\nu})
\overline{\sigma}_{\mu} u(p_N)\right] \nonumber \\
& & \times  \left[\frac{1}{(p_N-p_l^-)^2-m^2_W}+
\frac{g^{(l)}_L}{\cos ^2 \theta_W} \frac{1}{(p_N-p_{\nu})^2-m^2_Z}
\right] \eeqa In the low-energy limit the expression in parentheses
reduces to \beq \frac{-1}{m^2_W} \left( \frac{1}{2} + \sin ^2 \theta
_W \right) \eeq In this limit the rate is proportional to \beq
 \left( \frac{1}{2} + \sin ^2 \theta _W \right)^2 \simeq 0.53 ~.
 \label{llinterference} \eeq
To this must be added the rate for neutral current decays into
$l^-_R l^+_L$, which is non-interfering and proportional to $(\sin^2
\theta_W)^2 \simeq 0.05$, to obtain a total rate for the $l^-_J
l^+_J \nu_J$ channel which is proportional to $0.59$ in the low
energy limit. The result (\ref{llinterference}) is somewhat smaller
than that obtained by incorrectly incoherently adding the charged
and neutral current amplitudes for decays into left-handed charged
leptons, which gives $1 + (-1/2+\sin^2 \theta _W)^2 \simeq 1.07$.

The charged-conjugated decay $N_I\rightarrow l^-_J l^+_J
\overline{\nu}_J$ has an identical rate to (\ref{llinterference}).
It adds incoherently to the amplitudes above because it has a
distinguishable final state; one final state has a neutrino and the
other an anti-neutrino.}

\end{document}